\documentclass[traditabstract]{aa}
\usepackage{graphics}
\usepackage{amssymb, amsmath}
\usepackage{natbib}
\bibpunct{(}{)}{;}{a}{}{,}

\DeclareRobustCommand{\ion}[2]{%
\relax\ifmmode
\ifx\testbx\f@series
{\mathbf{#1\,\mathsc{#2}}}\else
{\mathrm{#1\,\mathsc{#2}}}\fi
\else\textup{#1\,{\mdseries\textsc{#2}}}%
\fi}

\def\apjl{ApJ}%
\def\apjs{ApJS}%
\def\ao{Appl.~Opt.}%
\def\aap{A\&A}%
\begin{document}

\title{A Detailed Study of the Radio--FIR Correlation in NGC\,6946 with Herschel-PACS/SPIRE from KINGFISH}
\titlerunning{ Radio--FIR correlation in NGC\,6946}

\author{F.\,S. Tabatabaei\inst{1}, E. Schinnerer\inst{1}, E.\,J. Murphy\inst{2},  R. Beck\inst{3}, B. Groves\inst{1}, S. Meidt\inst{1}, M. Krause\inst{3}, H-W.~Rix\inst{1}, K. Sandstrom\inst{1},  A.\,F. Crocker\inst{4}, M. Galametz\inst{5}, G. Helou\inst{6}, C.\,D. Wilson\inst{7}, R.~Kennicutt\inst{5}, D. Calzetti\inst{4}, B. Draine\inst{8}, G. Aniano\inst{8}, D. Dale\inst{9},  G. Dumas\inst{10}, C.\,W. Engelbracht\inst{11,12}, K.\,D. Gordon\inst{13}, J. Hinz\inst{11}, K.\, Kreckel\inst{1}, E. Montiel\inst{11}, H. Roussel\inst{14} }
\authorrunning{Tabatabaei et al.}

\institute{Max-Planck-Institut f\"ur Astronomie, K\"onigstuhl 17, 69117 Heidelberg, Germany
\and Observatories of the Carnegie Institution for Science,  Pasadena, CA 91101, USA
\and Max-Planck Institut f\"ur Radioastronomie, Auf dem H\"ugel 69, 53121 Bonn, Germany
\and Department of Astronomy, University of Massachusetts, Amherst, MA 01003, USA
\and Institute of Astronomy, University of Cambridge, Madingley Road, Cambridge CB3 0HA, UK
\and Infrared Processing and Analysis Center, MS 100-22, Pasadena, CA 91125, USA
\and Department of Physics \& Astronomy, McMaster University, Hamilton, Ontario L8S 4M1, Canada
\and Princeton University Observatory, Peyton Hall, Princeton, NJ 08544-1001, USA
\and Department of Physics \& Astronomy, University of Wyoming, Laramie, WY 82071, USA
\and Institut de RadioAstronomie Millim\'etrique, 38406 Grenoble, France
\and Steward Observatory, University of Arizona, 933 N. Cherry Ave., Tucson, AZ 85721, USA
\and Raytheon Company, 1151 E. Hermans Road, Tucson, AZ 85756, USA
\and Space Telescope Science Institute, 3700 San Martin Drive, Baltimore, MD 21218, USA
\and Institut d' Astrophysique de Paris, Universit\'e Pierre et Marie Curie (UPMC), CNRS (UMR 7095), 75014 Paris, France 
}

\offprints{F.\,S. Tabatabaei \\ taba@mpia.de}



\abstract{We derive the distribution of  the synchrotron spectral index across NGC\,6946 and investigate the correlation between the radio continuum (synchrotron) and far-infrared (FIR) emission using the KINGFISH Herschel PACS and SPIRE data.  The radio--FIR correlation is studied as a function of star formation rate, magnetic field strength, radiation field strength, and the total gas surface brightness.   
The synchrotron emission follows both star-forming regions and the so-called magnetic arms present in the inter-arm regions. The synchrotron spectral index is steepest along the magnetic arms ($\alpha_n \sim 1$), while it is flat in places of giant H{\sc ii} regions and in the center of the galaxy ($\alpha_n \sim 0.6-0.7$). The map of $\alpha_n$ provides an observational evidence for aging and energy loss of cosmic ray electrons propagating in the disk of the galaxy. 
Variations in the synchrotron--FIR correlation across the galaxy are shown to be a function of both star formation and magnetic fields.  We find that the synchrotron emission correlates better with cold rather than with warm dust emission, when the interstellar radiation field is the main heating source of dust.  The synchrotron--FIR correlation suggests a coupling between the magnetic field and the gas density.
NGC\,6946 shows a power-law behavior between the total (turbulent)  magnetic field  strength B and the star formation rate surface density $\Sigma_{\rm SFR}$ with an index of 0.14\,(0.16)$\pm0.01$.  This indicates an efficient production of the turbulent magnetic field with the increasing gas turbulence expected in actively star forming regions. Moreover, it is suggested that the B-$\Sigma_{\rm SFR}$ power law index is similar for the turbulent and the  total fields in normal
galaxies, while it is steeper for the turbulent than for the total fields in
galaxies interacting with the cluster environment. The scale-by-scale analysis of the synchrotron--FIR correlation indicates that the ISM affects the propagation of old/diffused cosmic ray electrons, resulting in a diffusion coefficient of $D_0=4.6\times 10^{28}$\,cm$^2$\,s$^{-1}$ for 2.2\,GeV CREs.  }  

\maketitle

\section{Introduction}
The correlation between the radio and far-infrared (FIR) emission of galaxies has
been shown to be largely invariant over more than 4 orders of magnitude in luminosity \citep[e.g. ][]{Yun} and out to a redshift of z$\sim$3 \citep[e.g.][]{Sargent}.  
%
%
This correlation is conventionally explained by  the idea that the FIR and radio emission are both being driven by the energy input from massive stars, and thus star formation. However, this connection is complicated by the
observation that the FIR emission consists of at least two components;
one heated directly by massive stars (i.e.~the `warm' dust component),
and one heated by the diffuse interstellar radiation field or ISRF (i.e.~the
`cold' dust component) \citep[see e.g.][]{Darine_07}, which includes the emission from the old stellar population \citep[e.g. ][]{Xu_90, Bendo}.

Similarly, the radio emission consists of  two main components; the thermal,
free-free emission and nonthermal synchrotron emission. A direct connection between the free-free emission  (of thermal electrons in H{\sc ii} regions) and young massive stars is expected \citep[e.g.][]{Osterbrock_60}.  Conversely, the connection between the synchrotron emission and massive stars is complicated by the convection and diffusion of cosmic ray electrons (CREs) from their place of birth (supernova remnants, SNRs) and by the magnetic fields that regulate the synchrotron emission in the interstellar medium (ISM). 

Hence, warm dust emission /thermal radio emission can be
directly associated with young stars and a correlation between warm dust and thermal radio  emission is  not surprising. On the other hand, the connection between cold dust emission /nonthermal synchrotron emission and massive stars (and thus star formation) is less clear.
A better correlation of the FIR with the thermal than the nonthermal radio emission has already been shown in the LMC, M\,31, and M\,33 \citep{Hughes_etal_06,Hoernes_etal_98,Tabatabaei_1_07}. 
The CREs experience various energy losses while interacting with matter and magnetic fields in the ISM,
causing the power law index of their energy distribution to vary.
Significant variation of the nonthermal spectral index was found in M\,33 with a flatter synchrotron
spectrum in regions of massive SF than in the inter-arm regions and the
outer disk \citep[][]{Tabatabaei_3_07}.  
\begin{table}
\begin{center}
\caption{General parameters adopted for NGC\,6946.}
\begin{tabular}{ l l } 
\hline
\hline
Position of nucleus    & RA\,=\,$20^{h}34^{m}52.3^{s}$      \\
  \,\,\,(J2000)  &  DEC\,=\,$60^{\circ}09\arcmin14\arcsec$\\
Position angle of major axis$^{1}$   & 242$^{\circ}$ \\
Inclination$^{1}$   & 38$^{\circ}$  (0$^{\circ}$=face on)\\
Distance$^{2}$   & 6.8\,Mpc$^3$\\
\hline
\noalign {\medskip}
\multicolumn{2}{l}{$^{1}$ \cite{Boomsma}}\\
\multicolumn{2}{l}{$^{2}$ \cite{Karachentsev} }\\
\multicolumn{2}{l}{$^{3}$ 1$\arcmin$=\,1.7\,kpc along major axis}\\
\end{tabular}
\end{center}
\end{table}

The critical dependence of the synchrotron emission on both magnetic fields and CRE propagation could cause the nonlinearity in the synchrotron-FIR correlation seen globally for galaxy samples \citep[e.g. ][]{Niklas_977}. Propagation of the CREs can also cause dissimilarities of the synchrotron and FIR morphologies particularly on small scales.  For instance, Galactic  SNRs do not seem to be well-correlated with the FIR emission \citep[e.g. ][]{Cohen}. Moreover, within a few 100\,pc of the star-forming Orion nebula, no correlation exists  \citep{Boulanger_88}. In nearby galaxies, a lack of correlation on small scales has been shown via detailed multi-scale analysis using wavelet transformation \citep[e.g. ][]{Hughes_etal_06,Dumas}. \cite{Murphy_06,Murphy_08} showed that recent massive SF could  reduce these dissimilarities due to generation of a new episode of  CREs, assuming that the FIR emission is attributed to  dust heating by the same stars.

Assuming the massive SF  is as the source of both FIR and synchrotron emission, \cite{Helou_93} and \cite{Niklas_97}    considered a  coupling between magnetic field strength and gas density as the reason for
the tight radio--FIR correlation in spite of the sensitive dependence of the synchrotron emission on the magnetic field. A modified version of this model was suggested by \cite{Hoernes_etal_98} to explain the correlation between the cold dust heated by the ISRF and the synchrotron emission, whose energy sources are independent. The scale  at which the correlation breaks down provides an important constraint on these models, explaining the scale where static pressure equilibrium between the gas and CREs/magnetic fields holds.

Within nearby galaxies, variations in the radio--FIR  correlation  have been shown to exist by several authors
\citep[e.g.][]{Gordon_04,Murphy_06,Hughes_etal_06,Murphy_08,Dumas} through a change in the $q$ ratio \citep[][see Sect.\,7.3 for definition]{Helou_etal_85} or in the fitted slope. Furthermore, the smallest scale  at which the radio--FIR correlation holds is not the same from one galaxy to another \citep{Hughes_etal_06,Tabatabaei_1_07,Dumas}.  

As the variations in the radio--FIR correlation are possibly due to a range of different conditions such as  the star formation rate (SFR), magnetic fields, CRE propagation, radiation field and heating sources of dust, this correlation  can be used as a tool to study the unknown interplay between the ISM components and SF. These are addressed in this paper through a detailed study of the radio--FIR correlation in NGC\,6946.

NGC\,6946 is one of the largest spiral galaxies  on the sky  at a distance of 6.8\,Mpc \citep{Karachentsev}. Its low inclination (38$^{\circ}$) makes it ideal for mapping various astrophysical properties across the galaxy (Table~1). NGC\,6946 shows a multiple spiral structure with an exceptionally bright arm in the north-east of the galaxy. Having several bright giant H{\sc ii} regions, this Sc (SABc)  galaxy has active SF  as well as strong magnetic fields as traced by linearly polarized observations \citep{Beck_96a}. The global star-formation rate is $\simeq$7.1\,M$_{\odot}$\,yr$^{-1}$   \citep[as listed by][ assuming a Kroupa IMF and a mass range of 0.1-100\,\,M$_{\odot}$]{Kennicutt_11}.  The dynamical mass of this galaxy is $\simeq 1.9\times10^{11}$\,M$_{\odot}$ \citep{Crosthwaite}. This galaxy harbors a mild starburst nucleus \citep[e.g. ][]{Ball} and there is no strong evidence for AGN activity \citep[e.g. ][]{Tsai}.
  
\cite{Frick_etal_01} presented the wavelet analysis of the radio and mid-infrared (ISOCAM LW3) emission in NGC\,6946. 
Here we study this correlation for dust emission at FIR wavelengths with Herschel-PACS and SPIRE from the KINGFISH project \citep[Key Insights on Nearby Galaxies: a Far-Infrared Survey with Herschel,][]{Kennicutt_11} and using various approaches.    

The paper is organized as follows. The relevant data sets are described
in Sect. 2. After deriving the maps of the free-free and synchrotron emissions (Sect.~3), we derive the distribution of the synchrotron spectral index in Sect.~4.  We map the magnetic field strength in Sect.~5.  The radio--FIR correlation is calculated using various approaches i.e. the q-method, classical pixel-by-pixel correlation, and as a function of spatial scale in Sect.~6. We further discuss the correlations versus magnetic fields, SFR, radiation field and gas density (Sect.~7).  Finally, we summarize the results in Sect.~8.
\begin{table*}
\begin{center}
\caption{Images of NGC\,6946 used in this study. }
\begin{tabular}{ l l l l} 
\hline

Wavelength & Resolution & rms noise & Telescope \\

\hline
20\,cm       &  $15\arcsec$ &  23\,$\mu$Jy/beam &VLA+Effelsberg $^{1}$\\
3.5\,cm      &  $15\arcsec$ &  50\,$\mu$Jy/beam  &VLA+Effelsberg$^{2}$\\
250\,$\mu$m  &  $18\arcsec$ & 0.7\,MJy\,sr$^{-1}$ &Herschel-SPIRE$^{3}$\\
160\,$\mu$m  &  $12\arcsec$ & 2.2\,MJy\,sr$^{-1}$ &Herschel-PACS$^{3}$\\
100\,$\mu$m  &  $8\arcsec$ & 5\,MJy\,sr$^{-1}$ &Herschel-PACS$^{3}$\\
70\,$\mu$m  &  $6\arcsec$ & 5\,MJy\,sr$^{-1}$  &Herschel-PACS$^{3}$\\
6563\AA{}\,(H$\alpha$)  &$1.5\arcsec$ & 0.06\,$\mu$erg\,\,s$^{-1}$\,cm$^{-2}$\,sr$^{-1}$ &  KPNO$^{4}$ \\
HI-21\,cm   &  $6\arcsec$ &  1.4\,Jy/beam\,m\,s$^{-1}$ &VLA $^{5}$\\
CO(2-1)     &  $13\arcsec$ & 0.06\,K\,km\,s$^{-1}$  &IRAM-30m $^{6}$\\
\hline
\noalign {\medskip}
\multicolumn{3}{l}{$^{1}$ \cite{Beck_91,Beck_07}}\\
\multicolumn{3}{l}{$^{2}$ \cite{Beck_07}}\\
\multicolumn{3}{l}{$^{3}$ \cite{Kennicutt_11}}\\
\multicolumn{3}{l}{$^{4}$ \cite{Ferguson}}\\
\multicolumn{3}{l}{$^{5}$ \cite{Walter_08}}\\
\multicolumn{3}{l}{$^{6}$ \cite{Leroy_08}}\\
\end{tabular}
\end{center}
\end{table*}
\section{Data}
Table 2 summarizes the data used in this work. 
NGC\,6946 was observed with the Herschel Space Observatory as part of the KINGFISH project
\citep[][]{Kennicutt_11}
and was described in detail in \cite{Dale_12} and \cite{Aniano_12}. 
Observations with the PACS instrument \citep{Poglitsch} were carried out at 70, 100, 160\,$\mu$m in the Scan-Map mode. The PACS images were reduced by the Scanamorphos data reduction
pipeline \citep[][]{Roussel_10,Roussel_12}, version 12.5. Scanamorphos version 12.5 includes the latest PACS calibration available and aims to preserve the low surface brightness diffuse emission. The 250$\mu$m map was observed with the SPIRE instrument \citep{Griffin} and  reduced using the HIPE version spire-5.0.1894. The data were subtracted for the sky as detailed in \cite{Aniano_12}.

The 70, 100, 160 and 250\,$\mu$m images were convolved from their native PSFs to a Gaussian PSF with 18$\arcsec$ FWHM using the kernels from \cite{Aniano_11} and resampled  to a common pixel size of 6$\arcsec$ ($\sim$170\,pc).

The radio continuum (RC) data at 3.5 and 20\,cm are presented in \cite{Beck_91} and \cite{Beck_07}. At 3.5\,cm, NGC\,6946 was observed with the 100-m Effelsberg telescope of the MPIfR\footnote{The 100--m telescope at Effelsberg is operated by the Max-Planck-Institut f\"ur Radioastronomie (MPIfR) on behalf of the Max--Planck--Gesellschaft.}. The 20\,cm data were obtained from observations with the Very Large Array (VLA\footnote{The VLA is a facility of the National Radio Astronomy Observatory. The NRAO is operated by Associated Universities, Inc., under contract with the National Science Foundation.}) corrected for missing short spacings using the Effelsberg data at 20\,cm. To trace the ordered magnetic field,  the linearly polarized intensity data at 6\,cm presented in \cite{Beck_96a} were used. The average degree of polarization is $\simeq$30\% for the entire galaxy. 

To investigate the connection between the neutral gas and the magnetic field,  we used the total gas (HI + H$_2$) mass surface density map which was derived using the VLA data of the HI-21\,cm line \citep[obtained as part of THINGS,][]{Walter_08} and the IRAM 30-m CO(2-1) data from the HERACLES survey  as detailed in \cite{Bigiel_08} and \cite{Leroy_08}.  

We used the H$\alpha$ map of NGC\,6946 observed with the KPNO 0.9 m in a 23$\arcmin\times23\arcmin$ field of view and with 0.69$\arcsec$ pixel$^{-1}$ (resolution of 1.5$\arcsec$), subtracted for the continuum \citep{Ferguson} and foreground stars.  The contribution from the [NII] line emission was subtracted following \cite{kennicutt_08}. The H$\alpha$ emission is corrected for attenuation by the Galactic cirrus using the extinction value given by \cite{Schlegel}. The H$\alpha$ map has a calibration uncertainty of $\simeq$20\%.

The radio and H$\alpha$ maps were smoothed to 18$\arcsec$ resolution (using a Gaussian kernel). All the maps were normalized to the same  grid, geometry, and size before comparison. 
\section{Thermal/nonthermal separation}
Constraints on the origin and propagation of cosmic rays can be achieved by studying the variation in the spectral index of the synchrotron emission across external galaxies.  To determine the variation in the nonthermal radio spectral index,  we separate the thermal and nonthermal components using a thermal radio tracer (TRT) approach in which one of the hydrogen recombination lines is used as a template for the free-free emission \citep[see e.g. ][]{Dickinson,Tabatabaei_3_07}. For NGC\,6946, we use the H$\alpha$ line emission which is the brightest recombination line data available.  Both the free-free and the H$\alpha$ emission are linearly proportional to the number of ionizing photons produced by massive stars, assuming that the H$\alpha$ emitting medium is optically thick
to ionizing Lyman photons \citep[][see also Sect.~3.2]{Osterbrock,Rubin}. However, the observed H$\alpha$ emission may suffer from extinction by dust which will lead to an underestimate of the free-free emission. Hence, following  \cite{Tabatabaei_3_07} and \cite{Tabatabaei_10}, we first investigate the dust content of NGC\,6946 in an attempt to de-redden the observed H$\alpha$ emission. Then we compare our de-reddening method with the one based on a combination of the total infrared  (TIR) and H$\alpha$ luminosity \citep{Kennicutt_09}.  



\subsection{De-reddening of the H$\alpha$ emission}
Following \cite{Draine7}, the interstellar dust heating has been modeled in NGC\,6946 \citep{Aniano_12} assuming a $\delta$-function in radiation field intensity, $U$, coupled with a power-law distribution $U_{\rm min} < U < U_{\rm max}$,
\begin{equation}
\begin{split}
dM_{\rm dust}/dU = {M_{\rm dust} \Big[(1 - \gamma)\,\delta(U - U{\rm min})} \\
 + \,\,  \gamma \,\frac{\alpha-1}{U^{1-\alpha}_{\rm min} - U^{1-\alpha}_{\rm max}}\,U^{-\alpha}\Big],
\end{split}
\end{equation}
where $U$ is normalized to the local Galactic interstellar radiation field, $M_{\rm dust}$ is the total dust mass, and $(1 - \gamma)$ is the portion of the dust heated by the diffuse interstellar radiation field defined by $U = U_{\rm min}$. The minimum and maximum interstellar radiation field intensities span $0.01 < U_{\rm min} < 30$ and  $3 < {\rm log}\, U_{\rm max} < 8$ \citep[see ][]{Dale_01,Dale_12}. Fitting this model to the dust SED covering the wavelength range between 3.5\,$\mu$m and 250\,$\mu$m, pixel-by-pixel, results in the 18$\arcsec$ maps (with 6$\arcsec$ pixels) of the dust mass surface density ($\sigma$), the distribution of the radiation fields (U), and  the total infrared (TIR) luminosity emitted by the dust.

The map of $\sigma$ is equivalent to a map of the dust optical depth at H$\alpha$ wavelengths given by $\tau_{\rm H\alpha}= \sigma\,   \kappa_{\rm H\alpha}$ where  $\kappa_{\rm H\alpha}$ is the dust opacity. Taking into account both absorption and scattering, $\kappa_{\rm H\alpha} = 2.187 \times 10^{4}$\,cm$^2$\,g$^{-1}$, assuming a Milky-Way value of the total/selective extinction ratio of R$_{\rm v}$=3.1 \citep{Weingartner}. 
The distribution of $\tau_{\rm H\alpha}$  over the disk of NGC\,6946 overlaid with contours of the H$\alpha$ emission is shown in Fig.~1.   We note that the dust mass from
modeling the SED at 18" resolution may be overestimated by $\sim$20\% due to the lack of
longer wavelength constraints \citep{Aniano_12}.

\begin{figure*}
\begin{center}
\resizebox{14cm}{!}{\includegraphics*{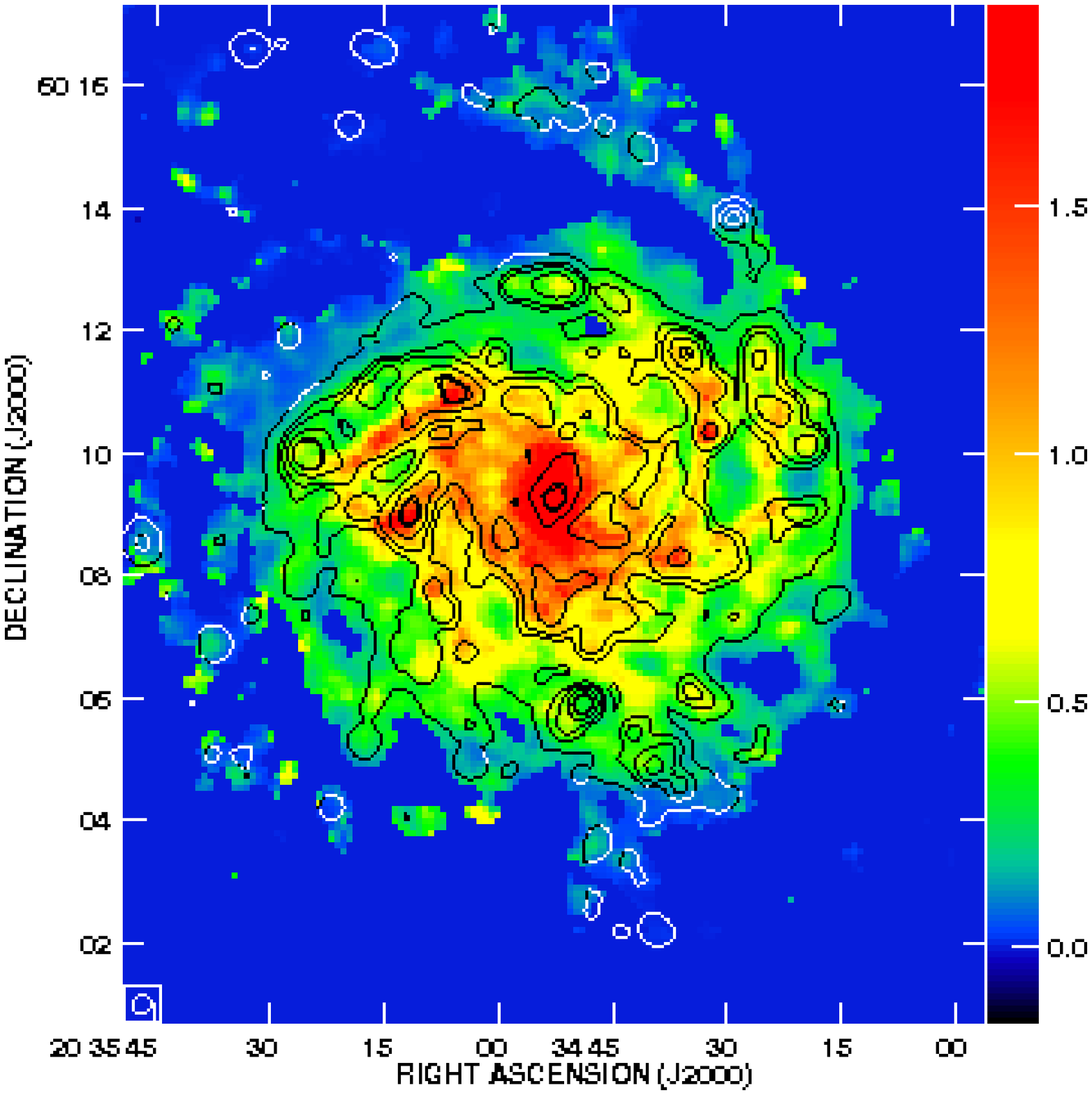}\includegraphics*{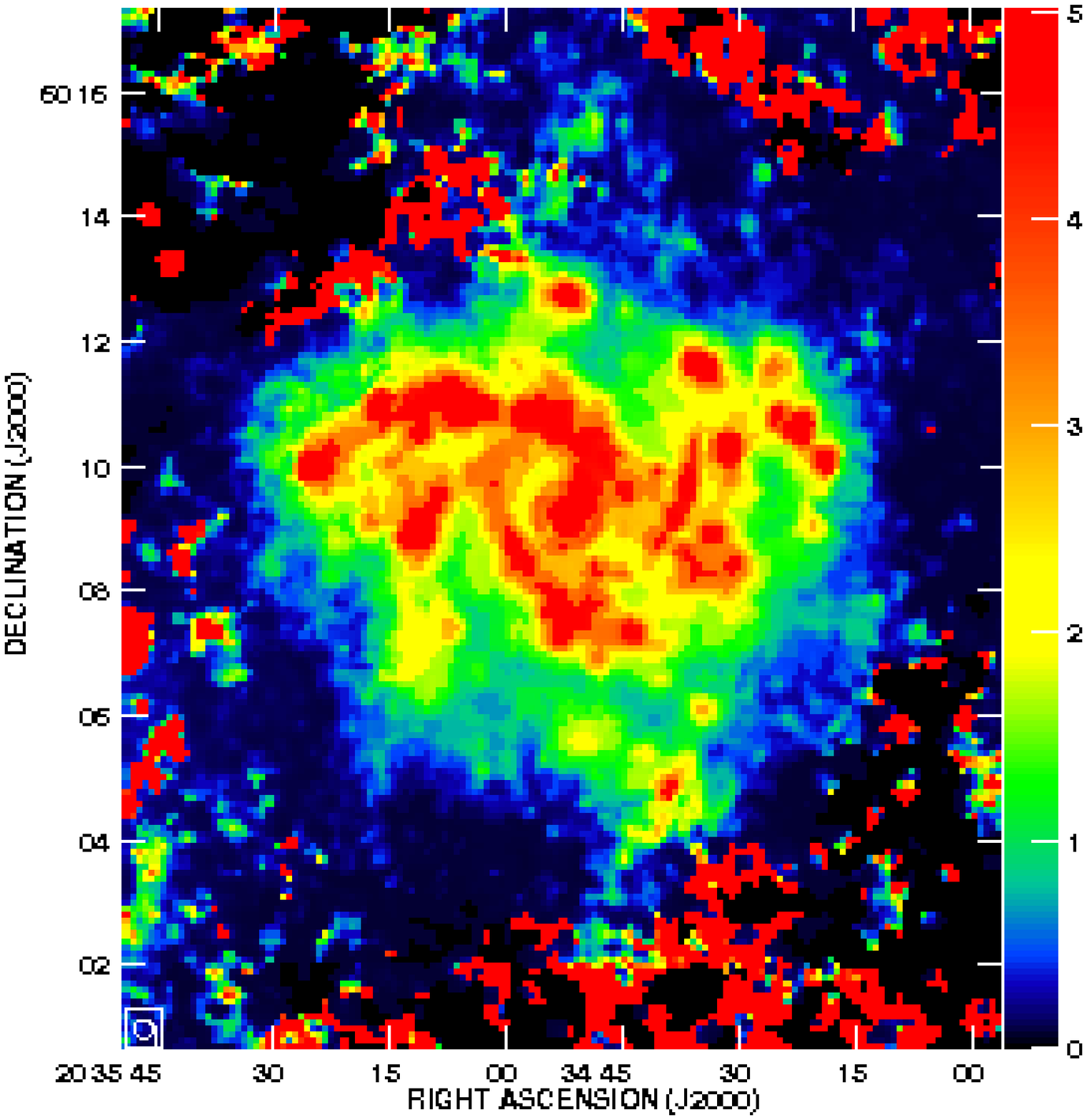}}
\caption[]{{\it Left:} Dust optical depth at H$\alpha$ wavelength $\tau_{{\rm H}{\alpha}}$ overlaid with contours of the H$\alpha$ emission from NGC\,6946. Contour levels are 1, 3, 6, 15, 30, 40\,$\mu$erg\,\,s$^{-1}$\,cm$^{-2}$\,sr$^{-1}$. The bar at the right shows the values of $\tau_{{\rm H}{\alpha}}$. {\it Right:} Radiation field $U$ map (mass-weighted mean starlight heating intensity) of NGC\,6946  in units of radiation field in solar neighborhood $U_{\odot}$ \citep{Aniano_12}, indicated by the bar at the right of the image. The beam area is shown in the lower left corners. }
\end{center}
\end{figure*}

Regions with considerable dust opacity ($\tau_{{\rm H}{\alpha} } >$ 0.7) follow narrow dust lanes along the spiral arms (e.g. the narrow lane in the inner edge of the bright optical arm) and are found mainly in the inner disk. 
High opacity dust is found in the center of the galaxy with $\tau_{{\rm H}{\alpha}}\simeq\,5$.  This corresponds to a silicate optical depth of $\tau_{9.7} \simeq$\,0.5 which is in agreement with \cite{Smith_07}. In the central $\sim$60\,pc which is much smaller than our resolution, much larger estimates of the extinction have been found using total gas masses \citep[][]{Schinnerer_6}.

Figure~1 also shows that the  H{\sc ii} complexes are dustier in the inner disk (with $\tau_{{\rm H}{\alpha} } \gtrsim$\,1.4) than in the outer parts (with $\tau_{{\rm H}{\alpha} } \lesssim$\,0.6) of NGC\,6946.    Across the galaxy, the mean value of  $\tau_{{\rm H}{\alpha} }$ is $0.43 \pm 0.04$ (median of $0.34 \pm 0.04$). Therefore, apart from the center, NGC\,6946 is almost transparent to photons with $\lambda \simeq 6563$\AA\ propagating towards us. 

The $\tau_{{\rm H}{\alpha} }$ derived can then be used to correct the H$\alpha$ emission for attenuation by dust, taking into account the effective fraction of dust actually absorbing the H$\alpha$ photons. Since these photons are usually emitted from sources {\it within} the galaxy, the total dust thickness $\tau_{{\rm H}\alpha}$ only provides an upper
limit.  Following \cite{Dickinson}, we set the effective thickness to $\tau_{\rm
eff} = f_d \times \tau_{{\rm H}\alpha}$ with $f_d$ being the dust fraction actually absorbing the H$\alpha$; the attenuation factor
for the H$\alpha$ flux is then $e^{-\tau_{\rm eff}}$. At our resolution of 18$\arcsec \simeq$530\,pc, one may assume that the H$\alpha$ emitting ionized gas is uniformly mixed with the dust, which would imply $f_d \simeq 0.5$. Considering the fact that the ionized gas has a larger extent than the dust, Dickinson et
al. (2003) found a smaller effective factor ($f_d=0.33$) based on a cosecant law modeling. We also adopt $f_d=0.33$ for NGC\,6946. We note that this choice barely influences the thermal fraction of the
radio emission, due to the small  $\tau_{{\rm H}{\alpha} }$ (Sect.~3.2).
%

Of course, it would be preferable not to use a uniform value $f_d$ for the whole galaxy, but one that is adapted to the geometry \citep[well mixed diffuse medium or shell-like in H{\sc ii} regions, ][]{Witt} and the dust column density. However, this would require specifying the location of the stellar sources and the absorbing dust along the line of sight and solving the radiative transfer problem with massive numerical computations which is far from our first order approximation. 



In another approach, assuming that the dust is mainly heated by the massive stars, we corrected the H$\alpha$ emission by combining it with the TIR (integrated dust luminosity in the 8-1000\,$\mu$m wavelength range): H$\alpha_{\rm corr}$=\,H$\alpha_{\rm obs}$ + 0.0024\, TIR \citep{Kennicutt_09}. 
Interestingly, this approach is linearly correlated with the de-reddening using $\tau_{\rm eff}$ (Fig.~2), with an offset of 0.19\,dex and a dispersion of 0.14\,dex.  Figure~2 shows that at the highest luminosities, the corrected H$\alpha$  values agree, corresponding to the calibration of the second approach
specifically to star-forming regions. Outside of these regions, the
H$\alpha$-TIR ratio approach overestimates the correction applied to the
observed H$\alpha$, probably because of contributions from other
dust-heating sources. Masking out the diffuse emission in the inter-arm regions and outer disk (i.e., considering only the spiral arms and SF regions), both the offset and dispersion reduces to 0.11\,dex and hence both methods agree within  the uncertainties ($\sim$ 20\%  due to calibration).  This likely indicates that the diffuse dust is not heated by the UV radiation of ionizing stars.

\begin{figure}
\begin{center}
\resizebox{\hsize}{!}{\includegraphics*{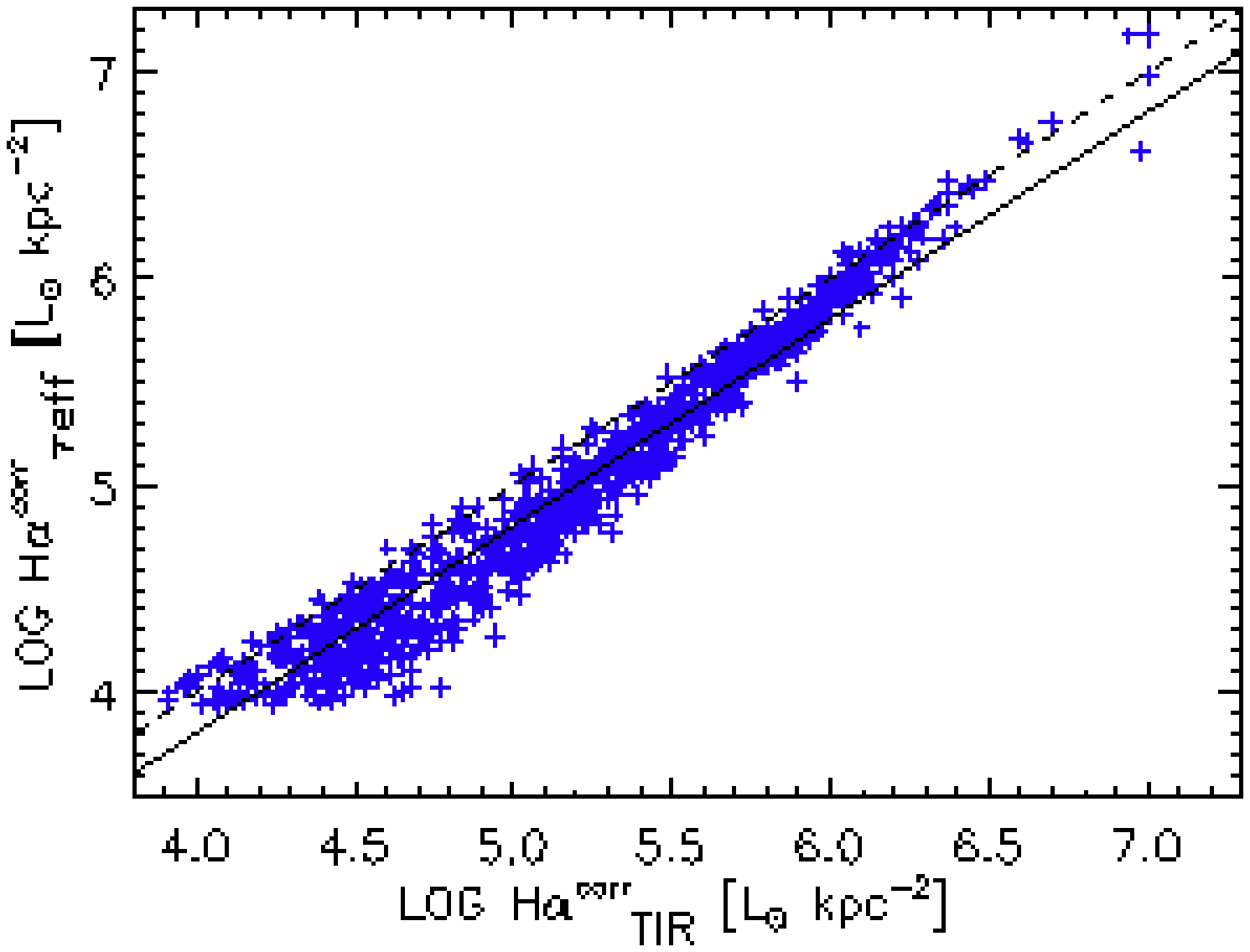}}
\caption[]{H$\alpha$ luminosity de-reddened based on $\tau_{\rm eff}$ versus de-reddening using the TIR luminosity. Also shown are the lines of 1:1 correspondence (dashed) and  Y=X - 0.19 (solid). }
\end{center}
\end{figure}

\begin{figure*}
\begin{center}
\resizebox{13cm}{!}{\includegraphics*{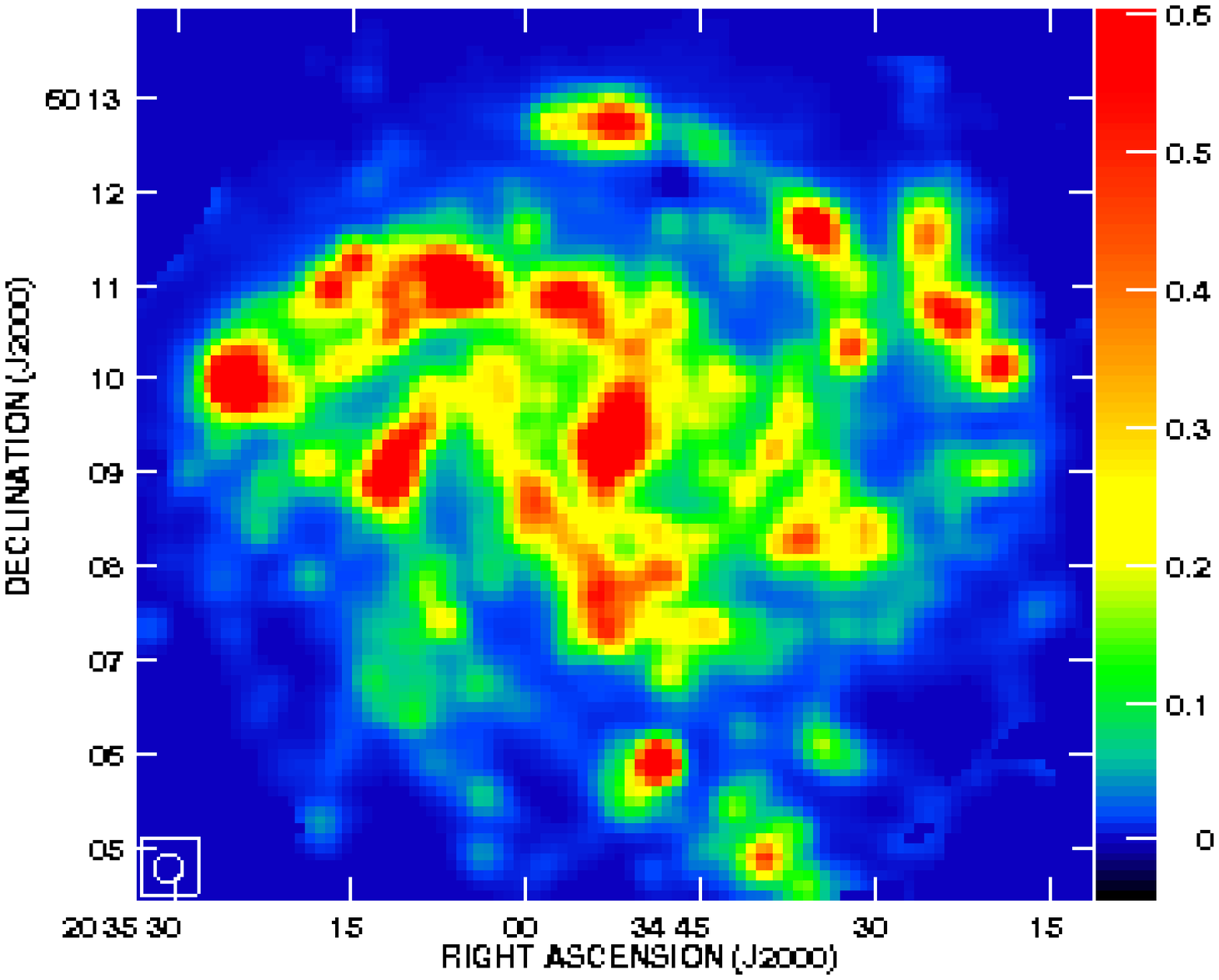}\includegraphics*{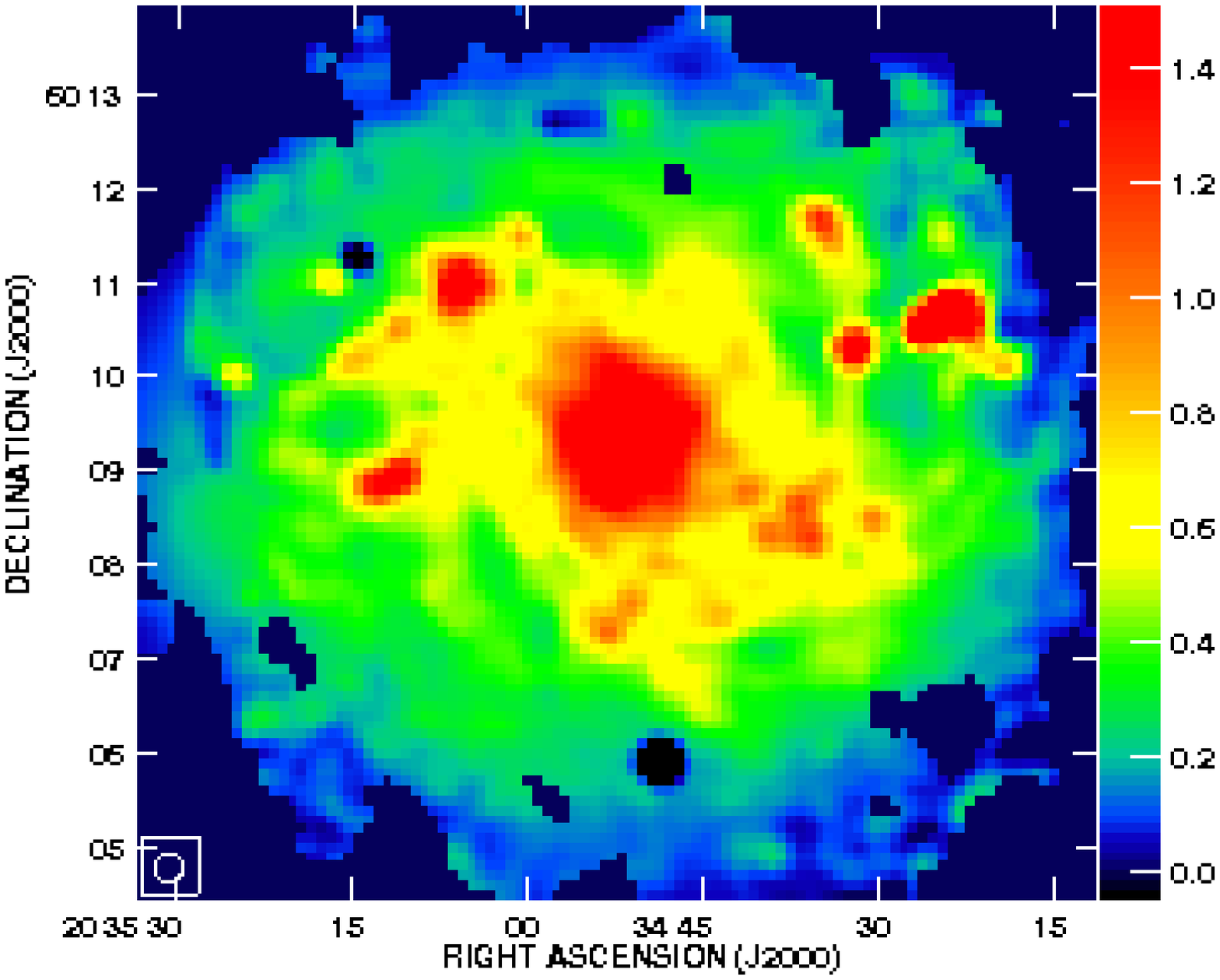}}
\resizebox{13cm}{!}{\includegraphics*{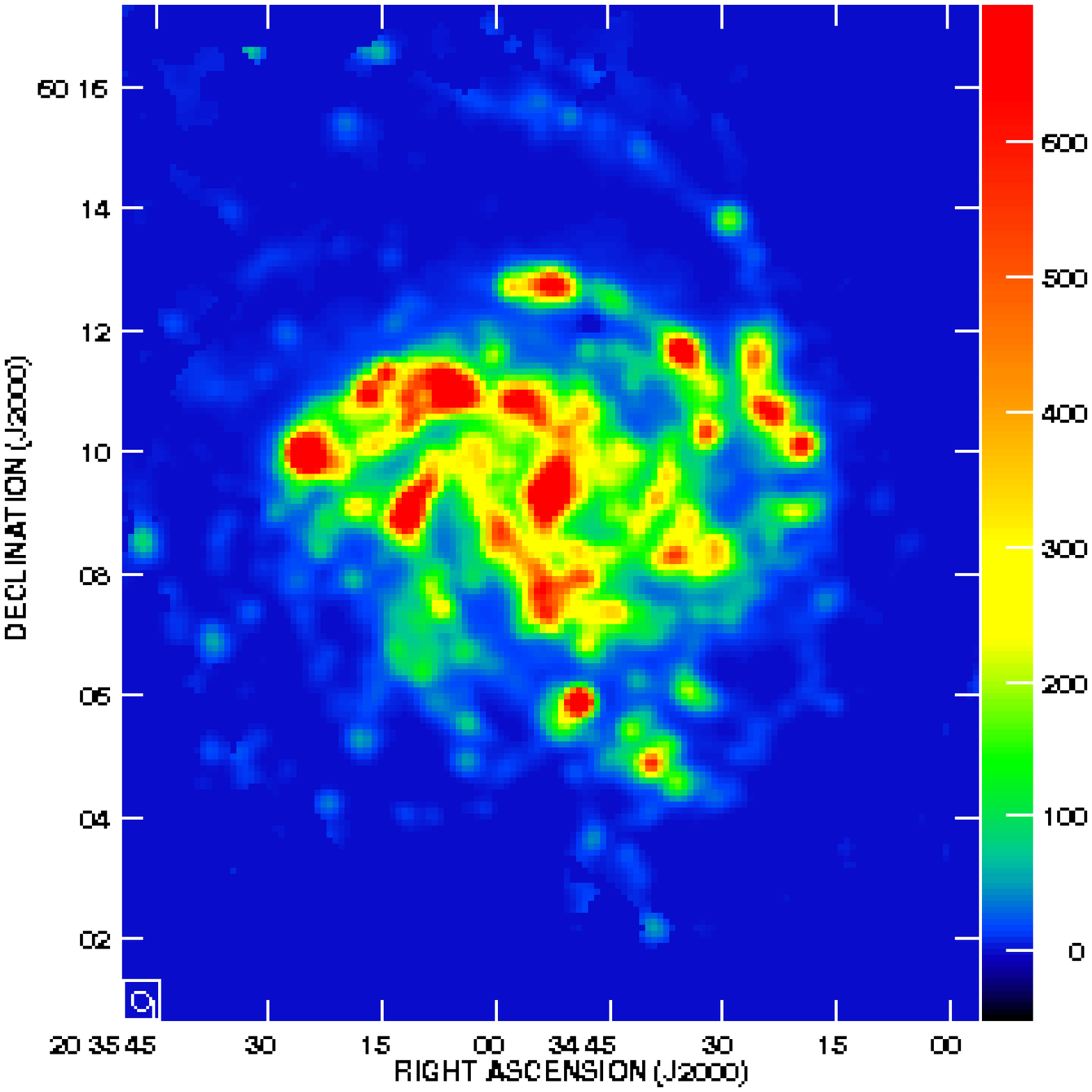}\includegraphics*{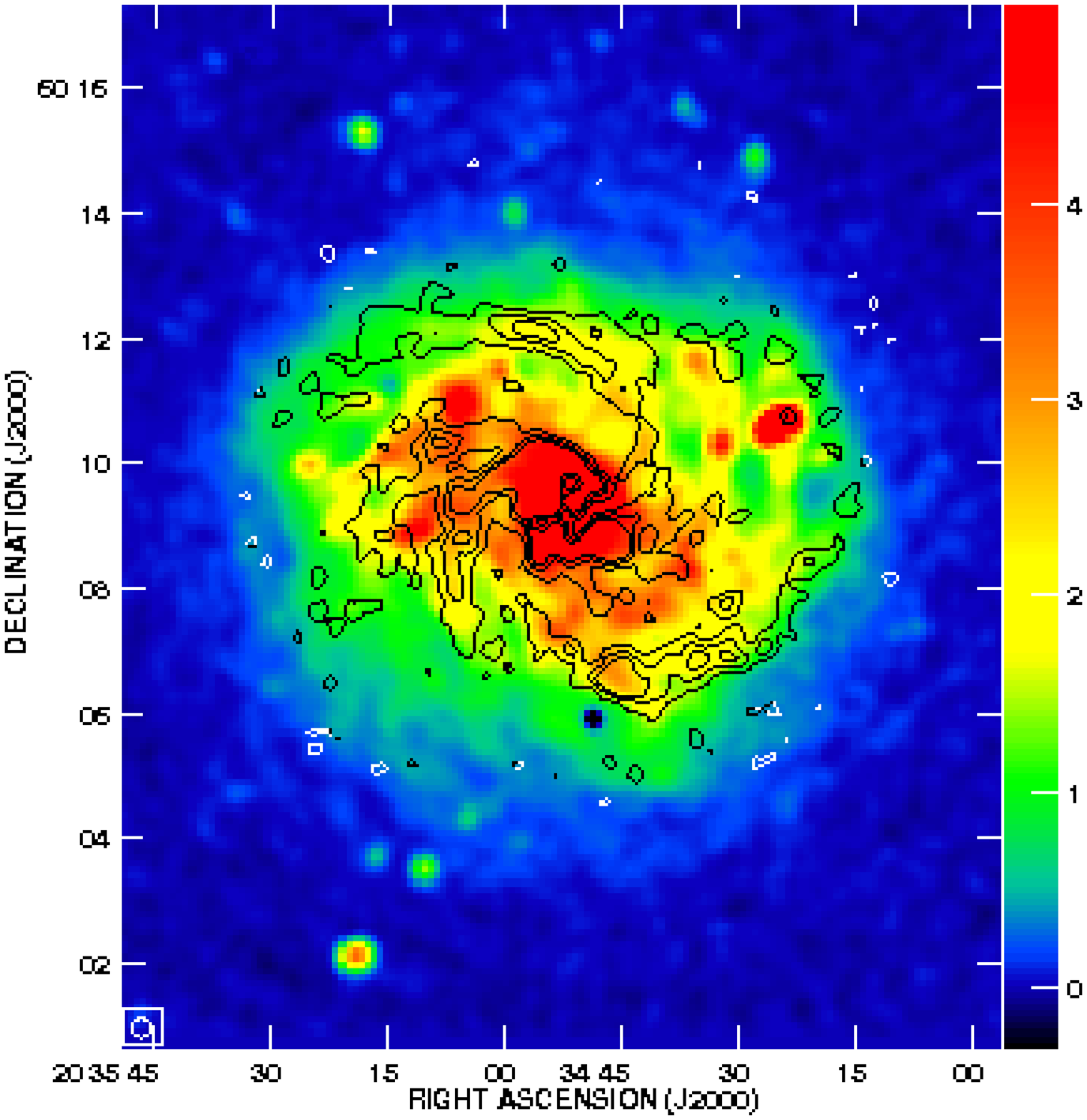}}
\caption[]{Free-free ({\it left}) and  synchrotron emission ({\it right}) from NGC\,6946 at 3.5\,cm ({\it top}) and 20\,cm ({\it bottom}). The angular resolution is 18$\arcsec$ (shown in the lower left corner of the panels) with a grid size of 6$\arcsec$.  The bars at the right of the images show the intensity values in mJy/beam. Note that the areas are not the same at 20\,cm and 3.5\,cm due to their different observed fields. At 20\,cm the synchrotron emission is overlaid with contours of linearly polarized intensity \citep[][]{Beck_96a}. The contour levels are 70, 120, 160\,$\mu$Jy/beam. }
\end{center}
\end{figure*}
\subsection{Distribution of the free-free and synchrotron emission}
Using the corrected H$\alpha$ emission from the first approach,  we derive the intrinsic H$\alpha$ intensity, $I_0$, according to $I = I_0 \,\,e^{-\tau_{\rm eff}}$.
Integration of the de-reddened H$\alpha$ map out to a radius of 11.9\,kpc (414$\arcsec$) yields a luminosity of $\rm L_{\rm H\alpha}= (3.46\pm0.05) \times 10^{41}$\,erg\,s$^{-1}$ that is higher than the foreground-corrected luminosity by $\simeq$\,20\%. A similar increase in the H$\alpha$ flux has been derived in other nearby galaxies, M\,33 \citep[$\simeq$\,13\%, ][]{Tabatabaei_3_07} and M31 \citep[$\simeq$\,30\%, ][]{Tabatabaei_10}. 
\cite{Dickinson} showed that the H$\alpha$ emitting medium in our Galaxy is
optically thick to ionizing Lyman photons \citep[case B, ][]{Osterbrock} not
only for H{\sc ii} regions ($\tau_{\rm Ly\alpha} \sim 10^3 - 10^{10}$) but also for faint H$\alpha$
features at intermediate and high Galactic latitudes ($\tau_{\rm Ly\alpha} \sim 1 - 30$). Assuming the same condition applies for NGC\,6946,  the emission measure (EM) follows from the H$\alpha$ intensity (in units of erg\,cm$^{-2}$\,s$^{-1}$ sr$^{-1}$) via the expression \citep{Valls} \,:

\begin{equation}
I_{{\rm H}\alpha} = 9.41 \times 10^{-8} T^{-1.017}_{e4}
10^{-\frac{0.029}{T_{e4}}} \, {\rm EM} \ ,
\end{equation}
where the electron temperature, $T_{e4}$, is in units of $10^4$\,K, and EM in
cm$^{-6}$ pc.  The emission measure is related to the continuum optical thickness, $\tau_c$, of the ionized gas by
\begin{equation}
\tau_c = 3.278 \times 10^{-7} a T_{e4}^{-1.35} \nu_{{\rm GHz}}^{-2.1} (1+0.08) \,
{\rm EM} \ ,
\end{equation}
with $a \simeq 1$ \citep{Dickinson}.  The factor (1\,+\,0.08) takes into account the contribution from  singly ionized He. The brightness temperature of the radio
continuum (free-free) emission, $T_b$, then follows from
\begin{equation}
T_b = T_e(1-e^{-\tau_c}) \ .
\end{equation}
Eq. (4) with Eqs.(2) and (3) gives:
%
\begin{equation}
\left \{ \begin{array}{ll}
{T_b=T_e(1-e^{-A\,I_{{\rm H}\alpha}})} ,  \\ 
A=3.763\,a\,\nu_{GHz}^{-2.1}\, T_{e4}^{-0.3}\, 10^{\frac{0.029}{T_{e4}}}.   %
\end{array} \right.
\end{equation}

Hence, the free-free emission can be derived separately at each radio wavelength.
The resulting distributions of the intensity of the free-free emission in mJy/beam at 3.5 and 20\,cm are shown in  (Fig.~3, left panels) for{\footnote{Smaller values of $T_e$ are reported from the measurements in the Milky Way \citep[e.g.][]{Haffner,Madsen}. Assuming $T_e=7000$\,K, the thermal fraction would decrease by about 23\%.  }  $T_e=10^4$\,K. Using a constant electron temperature is supported  by the  shallow metallicity gradient found in this galaxy \citep{Moustakas}. Subtracting the free-free emission from the observed radio continuum emission results in a map of the synchrotron emission (Fig.~3, right panels).  

The synchrotron maps exhibit diffuse emission extending to large radii indicating diffusion and propagation of the CREs. Strong synchrotron emission emerges from the galaxy center, giant star-forming regions, and spiral arms, which could be due to stronger magnetic fields and/or young-energetic CREs close to the star-forming regions. Interestingly, the so called 'magnetic arms' traced by the linearly polarized intensity \citep{Beck_96a} are clearly visible in the 20\,cm synchrotron map. The fact that they are less prominent at 3.5\,cm implies that these arms are filled by older and lower energetic CREs (see Sect~4.2).  The thermal free-free map, on the other hand, exhibits narrow spiral arms dominated by the star-forming regions. 

\begin{table}
\caption{Global radio continuum flux densities and thermal fractions in NGC\,6946.}
\begin{tabular}{ l l l l} 
\hline \hline
$\lambda$ &Observed &   Free-free & Thermal \\
(cm)&flux density   &  flux density& fraction \\
 &(mJy)& (mJy)& $\%$ \\
\hline
3.5 & 422\,$\pm$\,65&  78\,$\pm$\, 10 & 18.4\,$\pm$\,3.7\\
20 & 1444\,$\pm$\, 215 & 97\,$\pm$\, 13  & 6.7\,$\pm$\,1.3\\
\hline
\end{tabular}
\label{tab:flux}
\end{table}

Integrating the observed, synchrotron and free-free maps in the plane of the galaxy (i=38$^{\circ}$) around the center out to a radius of 324\arcsec (9.2\,kpc), we obtain the total flux densities and thermal fractions at 3.5 and 20\,cm (Table~3). The thermal fractions are about 18\% and  7\%  at 3.5 and 20\,cm, respectively. As mentioned before, we assumed a dust attenuation factor of $f_d=0.3$. For a uniform distribution of dust and ionized gas ($f_d=0.5$), the thermal fractions  increase to about 21\% and 8\% at 3.5 and 20\,cm, respectively.  
\begin{figure}
\begin{center}
\resizebox{\hsize}{!}{\includegraphics*{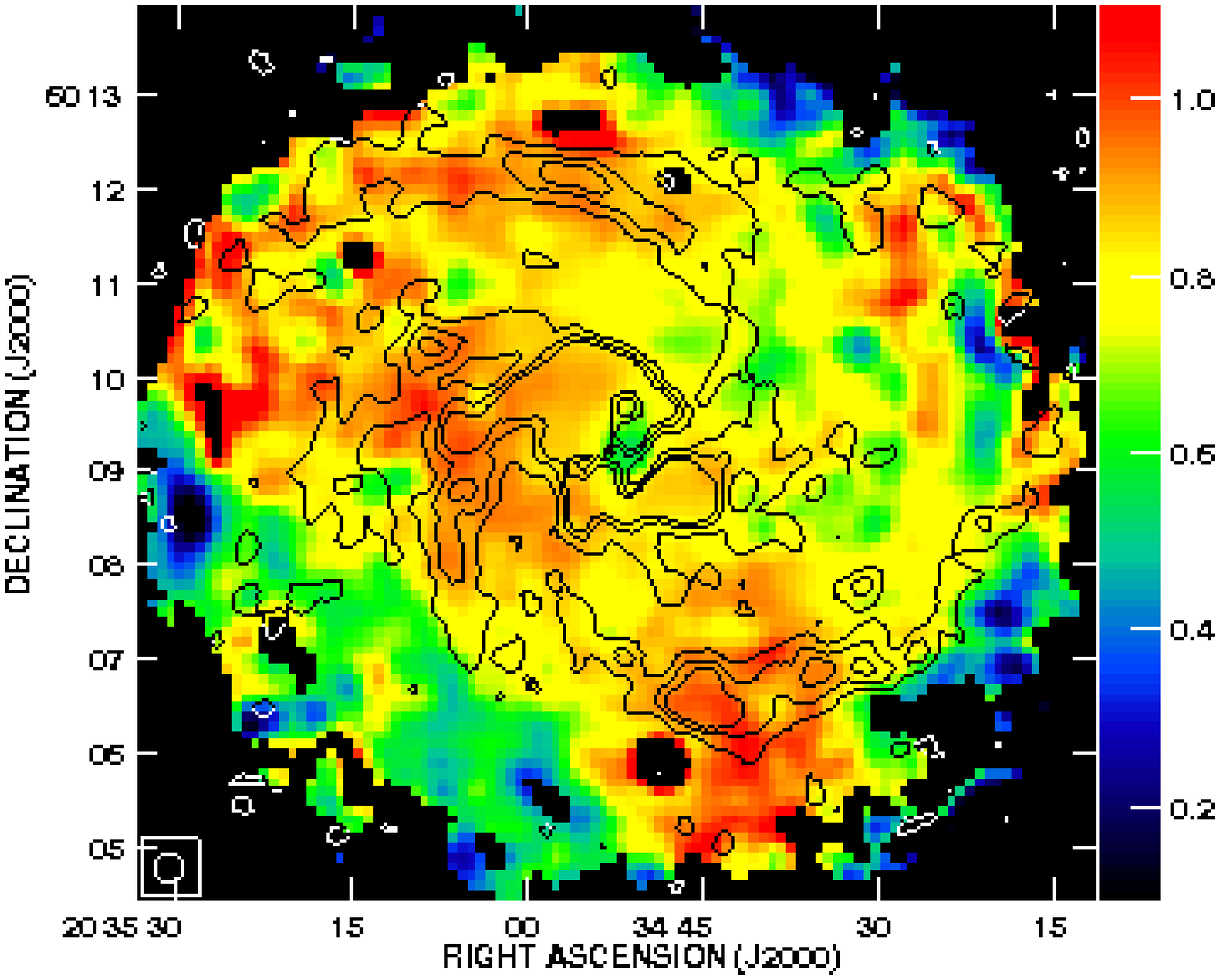}}
\resizebox{\hsize}{!}{\includegraphics*{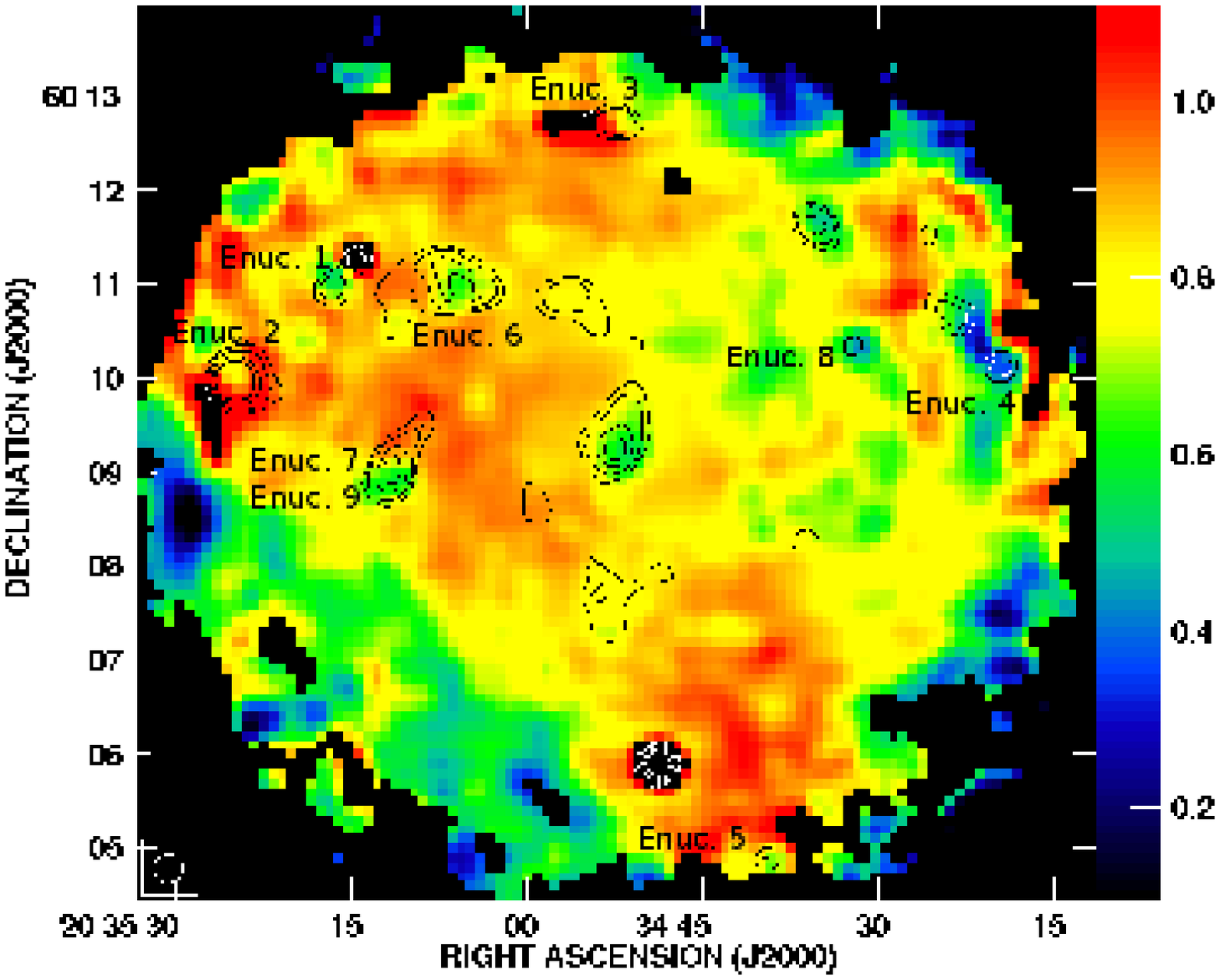}}
\caption[]{Synchrotron spectral index map  of NGC\,6946 overlaid with contours of  the linearly polarized emission at 6\,cm  ({\it top})  tracing the ordered magnetic field in the sky plane with the same levels as in Fig.\,3  and of bright H$\alpha$ sources ({\it bottom}). The H$\alpha$ contour levels are 18, 24, 35, and 47\% of the maximum intensity. The 9 giant H{\sc ii} regions are indicated (Table~4). The bars at the right of the images show the values of the synchrotron spectral index. }
\end{center}
\end{figure}

\begin{figure}
\begin{center}
\resizebox{\hsize}{!}{\includegraphics*{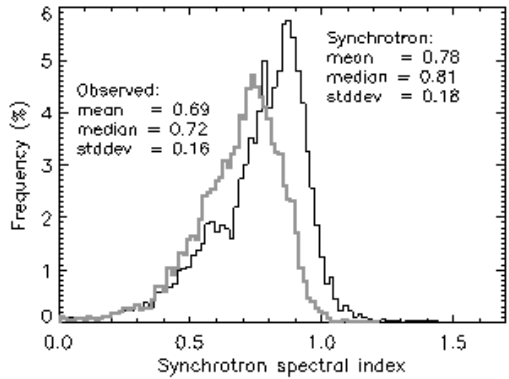}}
\caption[]{Histogram of the observed radio continuum (gray) and synchrotron spectral index (black) in NGC\,6946.}
\end{center}
\end{figure}
\section{Synchrotron spectral index} 
Using the nonthermal radio fluxes at 3.5 and 20\,cm, we obtained the spectral index of the nonthermal radio emission. This was only computed for pixels with flux densities of at least three times the rms noise at both frequencies. The synchrotron spectral index, $\alpha_n$, shows a smooth variation across the disk of NGC\,6946 (Fig.~4).  The greenish color in Fig.~4 corresponds to the synchrotron emission with flat spectrum ($\alpha_n\lesssim 0.7$), and the reddish color to the regions with a steep spectrum ($\alpha_n \gtrsim 0.9$), i.e. emission from lower-energy CREs. 
The flat spectrum is found in giant H{\sc ii} regions and the steep spectrum in the inter-arm regions, the eastern part of the central galaxy, and the south of the major axis where $\alpha_n= 1.0 \pm 0.1$. As shown in the histogram representation (Fig.~5), the median  value of $\alpha_n$  across the galaxy is 0.81 with a dispersion of 0.18. Figure 5 also presents a histogram of the `observed' spectral index $\alpha$ (obtained using the observed radio data at 3.5 and 20\,cm, i.e.,  contaminated by the free-free emission), for comparison, with $\alpha\simeq\,0.7$ on the average and with a dispersion of 0.16. 

\subsection{Synchrotron spectral index versus star formation} 
In the star-forming regions, the synchrotron spectrum is relatively flat with an average index of $\alpha_n$=\,0.65$\pm$0.10, the typical spectral index of young CREs in SF regions \citep[in supernova remnants, $\alpha_n$ is about 0.5 on the average and could even be flatter, see e.g.][and references therein]{Gordon_s_99, Reynolds_12,Longair}. Table~4 lists $\alpha_n$  in 9 giant H{\sc ii} regions annotated in Fig.\,4, \citep[called Enuc following the nomenclature and location presented in ][]{Murphy_10}. Enuc2 and Enuc3 have a relatively steep spectrum which could be due to an energy loss of CREs in a magneto-ionic  medium  along the line of sight seen at the position of these H{\sc ii} regions. This is possible for Enuc3 since this source is adjacent to the strong northern magnetic arm (see below). In the neighborhood of the Enuc2, however, no strong polarized emission is detected. The estimated spectral index of this source could be affected by an underestimate of the observed 3.5\,cm emission, since this  source sits on the edge of the 3.5\,cm observed field (primary beam). 

Considering synchrotron, inverse-Compton, ionization, and bremsstrahlung cooling mechanisms of CREs as well as using a prescription for the escape of these particles, \cite{Murphy_10} modeled the radio SEDs and found an $\alpha_n\simeq 0.8$ for the giant H{\sc ii} regions in Table~4, which is steeper than our finding and equals to the average $\alpha_n$ for the entire galaxy. One contributing factor to their steeper indices could be a result of
assuming a fixed ISM density for the extranuclear regions of 0.1\,cm$^{-3}$
(motivated by their beam size area of 30$\arcsec\sim$ 0.9\,kpc), leading to dominant synchrotron and IC losses rather than the ionization and bremsstrahlung cooling mechanisms of CREs (and hence a rather steep spectrum for these objects). In the present work, we determine $\alpha_n$ at a smaller beam size of $\sim$0.5\,kpc  \citep[the scale of the giant SF regions in NGC\,6946][]{Kennicutt_12} and without any assumption on the ISM density and/or cooling mechanism of CREs. On these scales, more energetic CREs and/or stronger magnetic fields close to the SF regions provide the synchrotron emission with a flatter $\alpha_n$ on the average. 

Table~4 also lists the thermal fractions  at 3.5\,cm and 20\,cm in these extranuclear H{\sc ii}
complexes. The thermal fractions at 3.5\,cm are higher than those at 20\,cm by a factor of about two (apart from Enuc4).  

\begin{table}
\caption{Thermal fractions and synchrotron spectral index of the giant H{\sc ii} regions shown in Fig.~4. }
\begin{tabular}{ l l l l l } 
\hline \hline
Object & $f_{\rm th}^{\rm 3.5cm}$ & $f_{\rm th}^{\rm 20cm}$  & \,\,\,\,\,$\alpha_n$ \\
 & (\%) & (\%) & \\
\hline
Enuc1 & 50\,$\pm$\,2 &26\,$\pm$\,1& 0.66\,$\pm$\,0.03\\
Enuc2 & 80\,$\pm$\,4 &41\,$\pm$\,2& 0.90\,$\pm$\,0.08\\
Enuc3 & 73\,$\pm$\,3 &34\,$\pm$\,2& 0.86\,$\pm$\,0.07\\
Enuc4 & 41\,$\pm$\,2 &26\,$\pm$\,1& 0.48\,$\pm$\,0.05\\
Enuc5 & 68\,$\pm$\,1 &34\,$\pm$\,1& 0.75\,$\pm$\,0.05\\
Enuc6 & 49\,$\pm$\,3 &23\,$\pm$\,2& 0.70\,$\pm$\,0.03\\
Enuc7 & 56\,$\pm$\,2 &27\,$\pm$\,2& 0.74\,$\pm$\,0.04\\
Enuc8 &  47\,$\pm$\,2 &20\,$\pm$\,1& 0.54\,$\pm$\,0.02\\
Enuc9 & 48\,$\pm$\,5 &23\,$\pm$\,3& 0.70\,$\pm$\,0.03\\
\hline
\end{tabular}
\label{tab:Fth}
\end{table}

\subsection{Synchrotron spectral index versus magnetic fields}
The magnetic fields in NGC\,6946 have been extensively studied by \cite{Beck_96a}, \cite{Rohde}, and \cite{Beck_07}. The contours in Fig.~4 show the linearly polarized intensity (PI) at 6.3\,cm which  determines the strength of the ordered magnetic field in the plane of the sky \citep[see e.g.][]{Tabatabaei_08}. Interestingly, there is a good correspondence between the steep synchrotron emission and the PI,  particularly along the northern magnetic arm \citep{Beck_96a} and also along the  strong ordered magnetic field in the central disk \citep[anisotropic turbulent magnetic field, ][]{Beck_07}. In these regions, the spectral index of $\alpha_n= 1.0 \pm 0.1$ indicates that CREs suffer strong synchrotron losses propagating along NGC\,6946's ordered magnetic field. 

Overall, the synchrotron spectral index map agrees with the energy loss theory of  relativistic electrons propagating away from their origin in star-forming regions in the ISM \citep[e.g. see Chapter 18 of][]{Longair,Biermann_01}. The difference in $\alpha_n$ in star-forming regions and along the magnetic arms in NGC\,6946 is similar to that predicted by \cite{Fletcher_11} for the spiral galaxy M51. 
\section{Maps of total and turbulent magnetic fields}
The strength of the total magnetic field B$_{\rm tot}$ can be derived from the total synchrotron intensity. Assuming equipartition between the energy densities of the magnetic field and cosmic rays ($\varepsilon_{CR}\,=\,\varepsilon_{\rm B_{\rm tot}} = {\rm B}_{\rm tot}^2/8\pi$):
\begin{eqnarray}
{\rm B}_{\rm tot}= C(\alpha_n, K, L) \big[I_{n} \big]^{\frac{1}{\alpha_n+3}},
\label{eq:Btoteq}
\end{eqnarray}
where $I_n$ is the nonthermal intensity and $C$ is a function of $\alpha_n$, $K$ the ratio between the number densities of cosmic ray protons and electrons, and $L$ the pathlength through the synchrotron emitting medium  \citep[see ][]{Beck_06,Tabatabaei_08}. Using the maps of $I_n$ and $\alpha_n$ obtained from the TRT method and assuming that the magnetic field is parallel to the plane of the galaxy (inclination of $i=38^{\circ}$ and position angle of the major axis of PA=242$^{\circ}$), B$_{\rm tot}$ is derived across the galaxy. In our calculations, we apply values of $K\,\simeq$\,100 \citep[][]{Beck_06} and  $L\,\simeq\,1\,{\rm kpc}/ {\rm cos}\,i$. Figure~6 shows strong B$_{\rm tot}$ in the central region of the galaxy, the arms and the star-forming regions.

\begin{figure}
\begin{center}
\resizebox{7cm}{!}{\includegraphics*{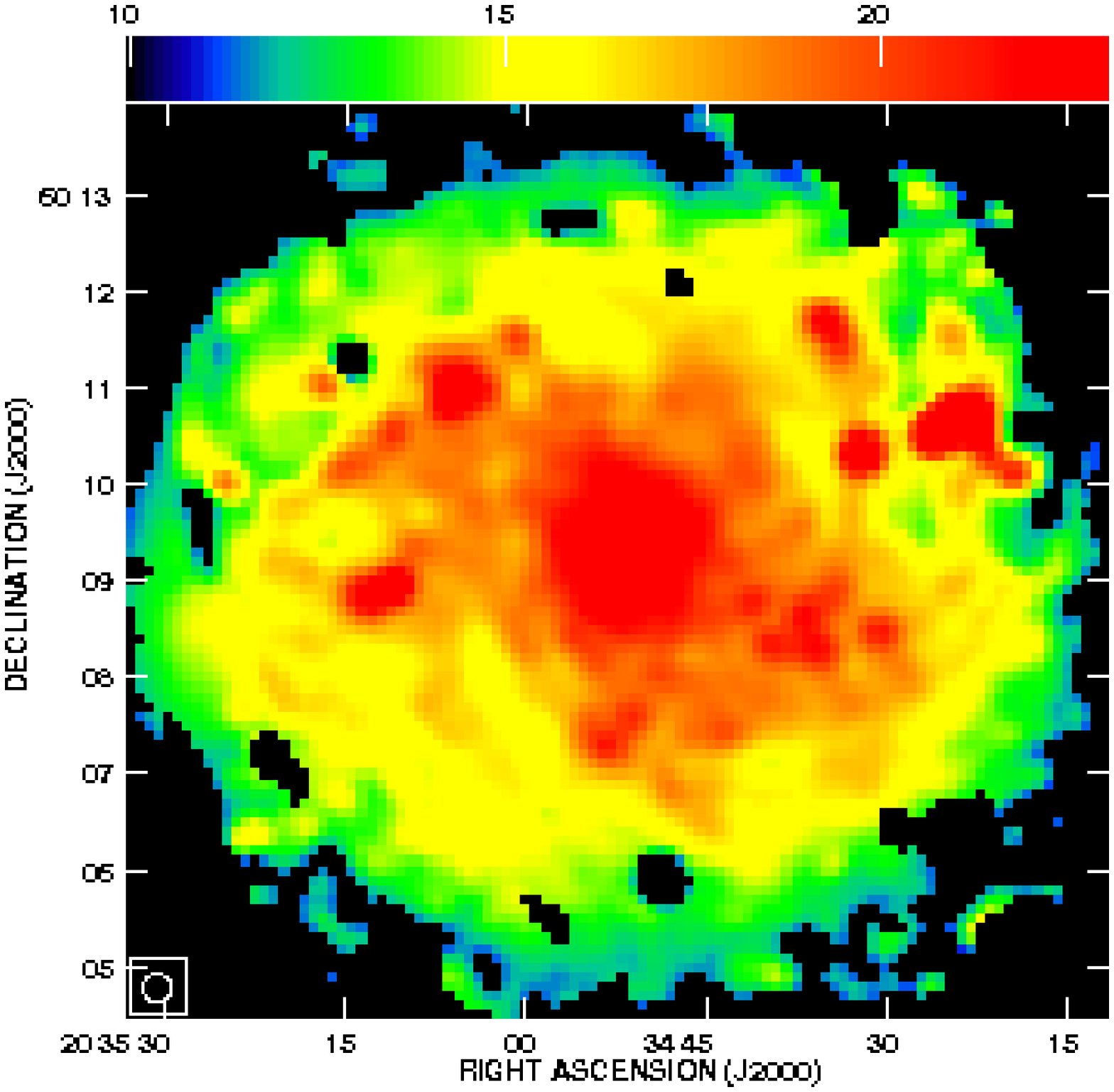}}
\resizebox{7cm}{!}{\includegraphics*{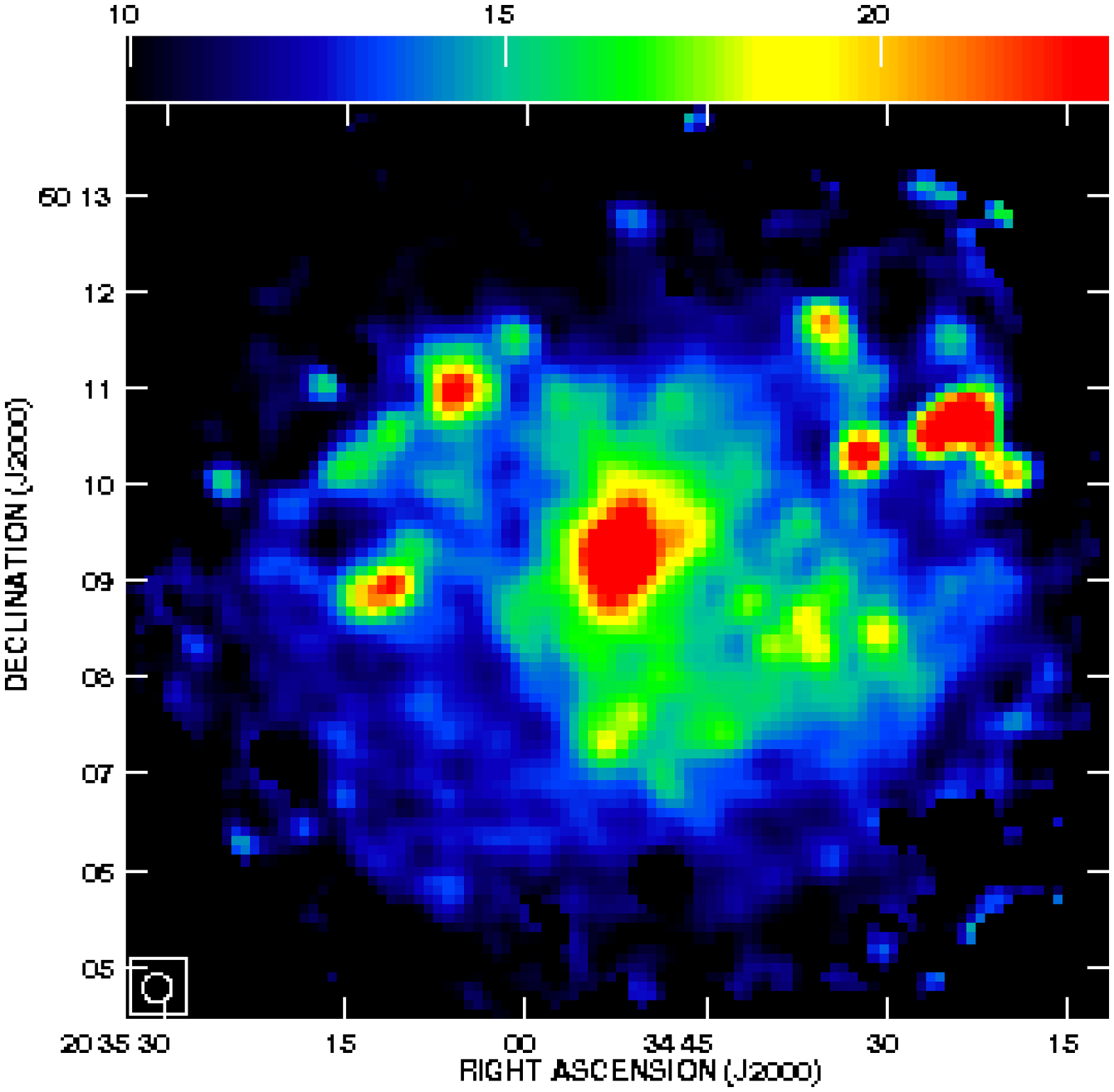}}
\caption[]{Strength of the total B$_{\rm tot}$ (top) and turbulent B$_{\rm tur}$ (bottom) magnetic fields in NGC\,6946. The bars at the top of each image show the magnetic field strength in $\mu$\,Gauss.}
\end{center}
\end{figure}

As a fraction of the polarized intensity  PI is related to the strength of the ordered magnetic field, and the nonthermal intensity $I_{n}$ to the total magnetic field in the plane of the sky, $I_n\,-\,$~(PI/0.75) gives the nonthermal emission due to the turbulent magnetic field B$_{\rm tur}$\footnote{In a completely ordered magnetic field, the maximum degree of  linear polarization  is about 0.75 \citep[e.g. ][]{Westfold}}. Using this intensity with Eq.~(\ref{eq:Btoteq}) yields the distribution of B$_{\rm tur}$ across the galaxy. Similar to B$_{\rm tot}$, B$_{\rm tur}$ is higher in places of star-forming regions (Fig.~6, bottom). We note that any ordered
field which is not resolved and depolarized within the beam would contribute to the turbulent magnetic field. In other words, we cannot distinguish between unresolved structures of the ordered field and truly turbulent fields for structures smaller than the beam.  

The magnetic field strength estimated using the `standard method' for thermal/nonthermal separation, i.e. assuming a fixed synchrotron spectral index (see Appendix A), is similar to the TRT estimate (within 10\% difference) in the spiral arms, and is underestimated by 35\% in the giant H{\sc ii} regions, and 45\% in the nucleus. This is because, in those regions,   $I_n$ is smaller and $\alpha_n$ is larger based on the standard method.

\section{Radio-FIR Correlation}
The radio correlations with both monochromatic and bolometric FIR observations are obtained using three different approaches: classical pixel-to-pixel (Pearson correlation), wavelet multi-scale \citep[][]{Frick_etal_01}, and the q-ratio method \citep{Helou_etal_85}.  Although the main goal of this paper is to investigate the synchrotron--FIR correlation, we also present  the correlations between the emissions of the thermal free-free and each of the monochromatic FIR bands.
\subsection{Classical correlation}
The pixel-to-pixel correlation is the simplest measure of the correlation between two images, $f_1(x,y)$
and $f_2(x,y)$, with the same angular resolution and the same number
of pixels using Pearson's linear correlation coefficient, $r_c$\footnote{Please note that using this method it is not possible to separate variations due to a change of the physical properties with scale e.g. propagation of CREs (see Sect.~6.2).}:
\begin{equation}
r_c = \frac{\Sigma(f_{1i} - \langle f_1 \rangle)(f_{2i} - \langle f_2 \rangle)}{\sqrt{\Sigma(f_{1i} - \langle f_1 \rangle)^2\,\Sigma({f_2i} - \langle f_2 \rangle)^2}} 
\end{equation}
When the two images are identical, $r_c$ = 1. The correlation coefficient is $r_c$=-1 for images that are perfectly anti-correlated. The formal error on the correlation
coefficient depends on the strength of the correlation and the number
of independent pixels, n,  in an image:
$\Delta r_{c}= \sqrt{1-r^{2}_{c}}/ \sqrt{n-2}.$

We calculated the correlations between both the radio free-free/synchrotron and the FIR 70, 100, 160, and 250$\mu$m maps, restricting the intensities to $>$\,3\,$\times$\, rms noise. We obtained sets of independent data points ($n$) i.e. a beam area overlap of $<20\%$, by choosing pixels spaced 
by more than one beamwidth. Since the correlated variables do not directly depend on each other, we fitted a power law to the bisector in each case \citep{Isobe,Hoernes_etal_98}. 
The Student's t-test is also calculated to indicate the statistical significance of the fit. For a number of independent points of $n\,>100$, the fit  is significant at the 3$\sigma$ level if $t>3$ \citep[e.g. ][]{Wall}.  Errors in the slope $b$ of the bisector are standard deviations (1\,$\sigma$).  


The results for both radio wavelengths are presented in Table~5. The calculated  Student's t-test values are large ($>33$) indicating that the fitted slopes are statistically significant. 
With coefficients of $r_c\geq 0.8$, good correlations hold between the FIR bands and observed radio (RC)/free-free emission. The FIR correlation coefficients with the synchrotron emission are slightly lower than those with the free-free emission. 

The synchrotron emission is slightly better correlated with the 250$\mu$m emission (as a proxy for cold dust) than  with the 70$\mu$m emission (as a proxy for warm dust). On the contrary, a free-free--cold/warm dust differentiation is only hinted, but not yet clear given the errors from the $r_c$ values. 

The free-free emission exhibits an almost linear correlation with the warmer dust emission at 70 and 100$\mu$m (with a slope of $b\simeq 0.9$). The correlation becomes more and more sub-linear with dust emission probed at 160 and 250$\mu$m. 

The synchrotron emission, on the other hand, tends to show a linear correlation with colder rather than  warmer dust (with a super-linear correlation). A similar trend is also seen between the observed RC and FIR bands.    
 
Super-linear radio--FIR correlations have been also found for samples of galaxies by \cite{Price} and \cite{Niklas_977}, which were attributed to the non-linearity of the synchrotron--FIR correlation and/or to the fact that colder dust may not be necessarily heated by the young massive stars.  A better synchrotron--cold than --warm dust correlation was also found in M\,31 \citep{Hoernes_etal_98} and in a sample of late-type galaxies \citep{Xu_94}. 
\begin{table*}
\begin{center}
\caption{Radio-IR relations on pixel-by-pixel basis in N6946.}
\begin{tabular}{ l l l l l l} 
\hline
  X    &    Y     &  \,\,\,\,\,\, $b$ &    $r_c$  & \, t & \,\,n\\
\hline 
\hline
RC(3.5)              & I$_{70}$  &     1.49$\pm$0.04 &  0.90$\pm$0.02 & 53.0 &662\\
                      &  I$_{100}$  &  1.49$\pm$0.03 &  0.92$\pm$0.01 & 61.8 & 696\\
                     &  I$_{160}$   &    1.20$\pm$0.03 &   0.88$\pm$0.01& 55.0 &883\\
                        &  I$_{250}$  &   1.03$\pm$0.03 &  0.89$\pm$0.02 & 56.3 &834\\
& & & & & \\
FF(3.5)             & I$_{70}$ &     0.92(0.95)$\pm$0.02 &  0.90(0.88)$\pm$0.02 & 52.5 &650 \\
                 &  I$_{100}$&  0.88(0.91)$\pm$0.02 &  0.90(0.89)$\pm$0.02 & 53.8&682\\
                &  I$_{160}$&    0.77(0.79)$\pm$0.01 &   0.88(0.88)$\pm$0.02 & 51.3&770\\
                &  I$_{250}$ &   0.67(0.69)$\pm$0.01 &  0.87(0.86)$\pm$0.02 & 48.2& 750\\
& & & & & \\
SYN(3.5)           & I$_{70}$ &     1.60(1.58)$\pm$0.04 &  0.80(0.82)$\pm$0.02 &33.8& 644\\
                 &  I$_{100}$&  1.59(1.57)$\pm$0.04 &  0.84(0.84)$\pm$0.02 &40.2 & 677\\
                &  I$_{160}$&    1.45(1.43)$\pm$0.03 &   0.86(0.87)$\pm$0.02 & 47.1& 784\\
                &  I$_{250}$ &   1.18(1.17)$\pm$0.03 &  0.86(0.87)$\pm$0.02 &46.4& 761\\
\hline
\hline
RC(20)             & I$_{70}$ &     1.52$\pm$0.03 &  0.87$\pm$0.02 &45.4  & 664\\
                  &  I$_{100}$  &  1.48$\pm$0.02 &  0.90$\pm$0.02 & 54.4 & 697\\
                   &  I$_{160}$&    1.16$\pm$0.02 &   0.92$\pm$0.01& 70.1 &911\\
                   &  I$_{250}$ &   1.02$\pm$0.01 &  0.92$\pm$0.01 & 68.3 &850 \\
& & & & & \\
FF(20)               & I$_{70}$ &     0.92$\pm$0.02 &  0.90$\pm$0.02 & 50.3 &656 \\
                 &  I$_{100}$&  0.87$\pm$0.02 &  0.90$\pm$0.02 & 52.5 &688\\
                &  I$_{160}$&    0.79$\pm$0.01 &   0.88$\pm$0.02 & 52.3&800\\
                &  I$_{250}$ &   0.68$\pm$0.01 &  0.87$\pm$0.02 & 49.3& 793\\
& & & & & \\
SYN(20)            & I$_{70}$ &     1.53$\pm$0.03 &  0.83$\pm$0.02 &38.2& 661\\
                 &  I$_{100}$&  1.48$\pm$0.03 &  0.87$\pm$0.02 &65.8& 695\\
                &  I$_{160}$&    1.19$\pm$0.02 &   0.91$\pm$0.01 & 65.7& 900\\
                &  I$_{250}$ &   1.04$\pm$0.02 &  0.90$\pm$0.01 &60.0&846\\
\hline
\end{tabular}
\tablefoot{Pearson correlation coefficients $r_c$ and bisector slopes $b$, log(Y)$\propto b$\,log(X), between the  total RC, Free-Free (FF), synchrotron (SYN) from the TRT method and the far-infrared PACS and SPIRE bands. The values in parenthesis show the corresponding correlations when the observed H$\alpha$ emission is directly used as the free-free tracer. Ordinary least-squares fits of bisector log(Y)$\sim~b\,log(X)$ are given through n pairs of (logX, logY), where n is the number of independent points (Isobe et al. 1990); $t$ is the  Student's t-test.}
\label{table:RIC3}
\end{center}
\end{table*}

One possible issue is that our use of the dust mass to de-redden the H$\alpha$ emission, and thus the free-free emission, is somehow influencing the correlations. To test this, we re-derive the free-free emission using the observed H$\alpha$ emission (not corrected for extinction) and re-visit the thermal/nonthermal correlations with the FIR bands. The results are given in parenthesis in  Table~5. The differences are less than 4\% and within the errors.

The synchrotron emission based on the standard method also shows a decrease of the slope with increasing  FIR wavelength (Table~6), similar to the TRT based study. However, the expected better linearity of the free-free--warmer dust is not seen using the standard thermal/nonthermal separation method.  As shown in Sect.~5, in our resolved study, this method results in an excess of the free-free diffuse emission in the inter-arm regions where there is no warm dust and TIR counterparts. This is most probably caused by neglecting  variations of $\alpha_n$ locally across the galaxy, since in global studies, the standard separation method leads to a linear thermal radio--FIR correlation \citep[e.g. ][]{Niklass}. Based on the standard method, the separated RC components are not as tightly correlated with the FIR bands as the observed RC--FIR correlations. 
      
\begin{table}
\begin{center}
\caption{Same as Table 5 for the 3.5\,cm Free-Free (FF)/synchrotron (SYN) emission separated using the standard method (see Appendix~A.). }
\begin{tabular}{ l l l l l l} 
\hline
  X    &    Y     &  \,\,\,\,\,\, $b$ &    $r_c$  & \, t & \,\,n\\
\hline 
\hline
FF(3.5)               & I$_{70}$ &     1.17$\pm$0.03 &  0.78$\pm$0.03 & 28.4 &508 \\
                 &  I$_{100}$&  1.13$\pm$0.03 &  0.75$\pm$0.03 & 25.5&518\\
                &  I$_{160}$&    1.03$\pm$0.03 &   0.72$\pm$0.03 & 24.3&543\\
                &  I$_{250}$ &   0.91$\pm$0.03 &  0.72$\pm$0.03 & 23.8& 540\\
& & & & & \\
SYN(3.5)            & I$_{70}$ &     1.60$\pm$0.05 &  0.77$\pm$0.03 &29.7& 594\\
                 &  I$_{100}$&  1.51$\pm$0.04 &  0.79$\pm$0.02 &33.9& 605\\
                &  I$_{160}$&    1.28$\pm$0.03 &   0.84$\pm$0.02 & 38.1& 626\\
                &  I$_{250}$ &   1.13$\pm$0.03 &  0.83$\pm$0.02 &37.5&622\\

\hline
\end{tabular}
\end{center}
\end{table}

\subsection{Multi-scale correlation}
Classical cross-correlations contain all scales that exist in a distribution. 
For example, the high-intensity points represent high-emission peaks on small spatial scales belonging to bright sources,  whereas low-intensity points represent weak diffuse emission  with a large-scale distribution, typically. 
However, such a correlation can be misleading when a bright, extended central region or an extended disk exists in the galactic image. In this case, the classical correlation is dominated by the large-scale
structure, e.g. the disc of the galaxy, while the (more interesting) correlation on the scale of the spiral arms can be much worse. The  classical correlation provides little information in the case of an anti-correlation on a specific scale \citep{Frick_etal_01}. Using the `wavelet cross--correlation' introduced by \cite{Frick_etal_01}, one can calculate the correlation between  different emissions as a function of the angular scale of the emitting regions. 
The wavelet coefficients of a 2D continuous wavelet transform are given by:
\begin{equation}
W(a,{\bf x})=\frac{1}{a^{\kappa}} \int_{-\infty}^{+\infty}  f(\bf x')\psi^{\ast}(\frac{{\bf x'-x}}{{\it a}}){\it d}{\bf x'},
\end{equation}
\noindent
where $f({\bf x})$ is a two--dimensional function (an image), $\psi({\bf x})$ is the analyzing wavelet, the symbol $^{\ast}$ denotes the complex conjugate, ${\bf x} = (x,y)$ defines the position of the wavelet  and {\it a} defines its scale.  $\kappa$ is a normalization parameter
(we use $\kappa=2$, which is equivalent to energy normalization). Following \cite{Frick_etal_01} and, e.g., \cite{Laine}, we use the `Pet Hat ' function as the  analyzing wavelet $\psi({\bf x})$ to have both a good scale resolution and a good spatial resolution.
%
The wavelet cross-correlation coefficient at scale {\it a} is defined as
\begin{equation}
r_{w}(a)=\frac{\int \int W_{1}(a,{\bf x})~W^{\ast}_{2}(a,{\bf x}) d{\bf x}}{[M_{1}(a)M_{2}(a)]^{1/2}},
\end{equation}
where the subscripts refer to two images of the same size and linear resolution and $M(a)= \int_{-\infty}^{+\infty} \int_{-\infty}^{+\infty}  \vert W(a,{\bf x})\vert ^{2} d\bf x$ is the wavelet equivalent of the power spectrum in Fourier space.   
The value of $r_{w}$  
varies between -1 (total anti-correlation) and +1 (total correlation). The correlation coefficient of  $|r_{w}|=0.5$ is translated as a marginal value for acceptance of the correlation between the structures at a given scale.
%
The error in $r_{w}$ is estimated by the degree of correlation and the number of independent points (n) as 
$\Delta r_{w}(a)= \sqrt{1-r^{2}_{w}}/ \sqrt{n-2}$,
where $ n\,\sim\,L_x L_y / a^2,  $ and $(L_x, L_y)$ are
the map sizes in $(x,y)$. Thus, towards larger scales, $ n$ decreases and the errors increase.

To prevent a strong influence of the nucleus on the wavelet analysis, the central 2\,kpc was subtracted from the PACS and SPIRE maps as well as the radio maps of NGC6946. Then, the maps were decomposed using Eq.~(8) into 12 angular scales ({\it a}) between 32\arcsec ($\sim$0.9\,kpc) and 313\arcsec ($\sim$9\,kpc)\footnote{The wavelet analysis was performed using the code described by \cite{Dumas}.}. The derived wavelet coefficients $W(a,{\bf x})$ of the  radio and FIR maps were then used in Eq.~(9) in order to derive the radio--FIR correlation coefficient $r_{w}$ on the 12 scales of decomposition. Figure~7 shows $r_{w}$ between the maps of the FIR bands and the 20\,cm synchrotron, free-free, and observed RC versus spatial scale $a$. 

The total RC--FIR and the synchrotron--FIR correlations are  higher for scales\,$>4-5$\,kpc, the scale of the diffuse central disk. 
On smaller scales the synchrotron--FIR correlations fluctuate  (within the errors) close to the threshold value $r_{w}$=0.5.  

On the smallest scale of 0.9\,kpc, corresponding to the width of the complexes of giant molecular clouds (GMCs) and star formation within spiral arms \citep[typical width of spiral arms in the galaxy\,$\sim 1.6-2$\,kpc, ][]{Frick_etal_01}, the synchrotron--colder dust correlation (e.g. 250$\mu$m, $r_w=0.53\pm$0.02) is slightly better than the synchrotron--warmer dust one (70$\mu$m, $r_w=0.45\pm$0.02). Right after a small maximum on scales of spiral arms ($\sim 1.6-2$\,kpc), a small minimum in the RC/synchrotron--FIR correlation occurs at $a\sim 2.5$\,kpc, the scale covering the width of the spiral arms plus diffuse emission in between the arms. Hence, the minimum could be due to the diffuse nature of the nonthermal emission caused by the propagation of CREs. The existence of the 'magnetic arms`  where there is no significant FIR emission, could also cause the general weaker correlation on scales of their widths, i.e.,  $<$\,3\,kpc (see Sect.~7.7).

The FIR bands are better correlated with free-free than with synchrotron emission on $a\leq\,3$\,kpc, as expected (note that this difference is not that obvious in the classical correlation at 20\,cm). Moreover correlations show a split in terms of the FIR band, with warmer dust having a better correlation than  colder dust. Such a split is again not obtained from the classical correlation. This occurs because the classical correlation is biased towards large scales (due to the bright disk of NGC\,6946), where the cold/warm dust split is less pronounced (or disappears in the free-free--FIR plot , Fig.~7 middle).


\begin{figure}
\begin{center}
\resizebox{\hsize}{!}{\includegraphics*{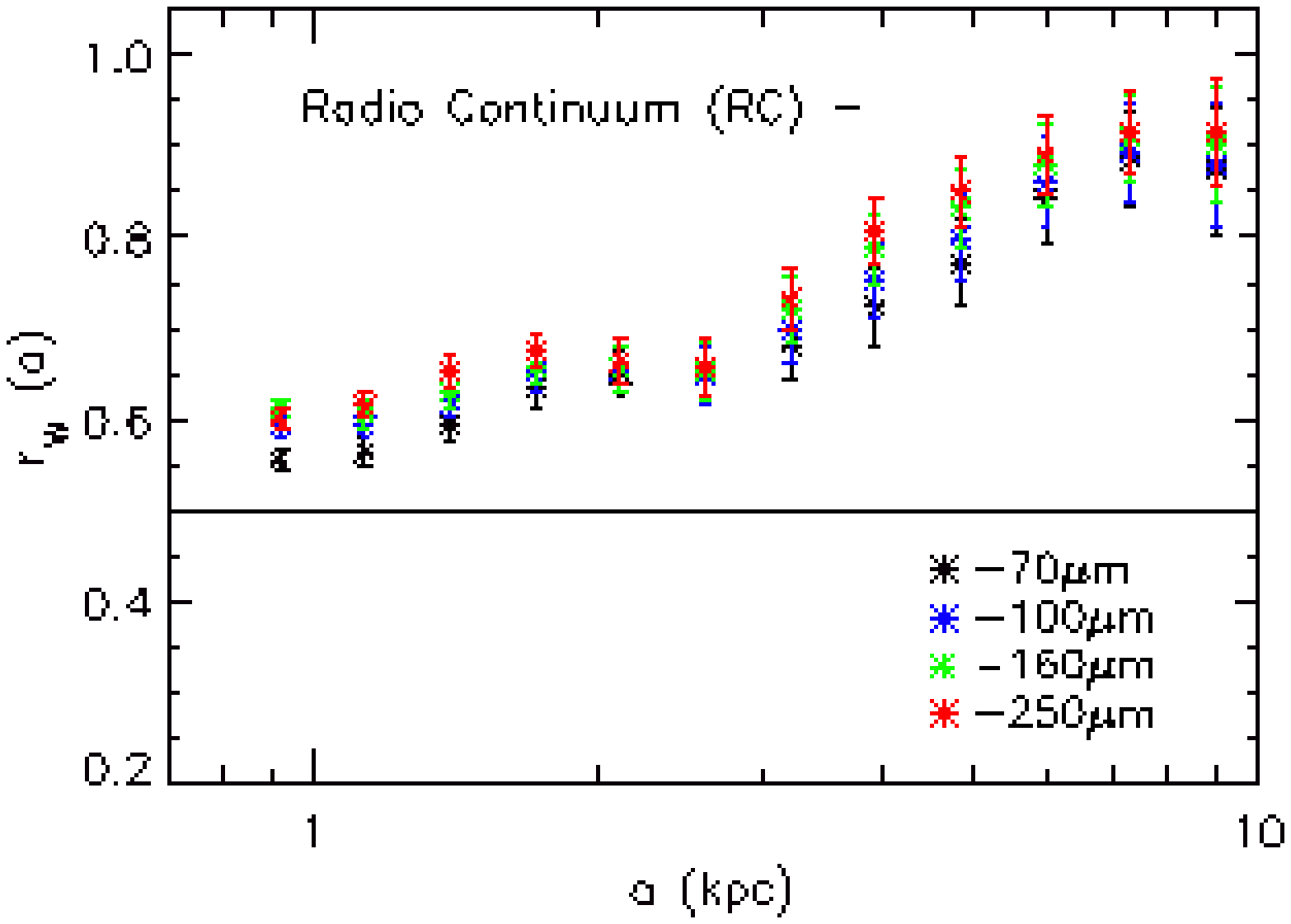}}
\resizebox{\hsize}{!}{\includegraphics*{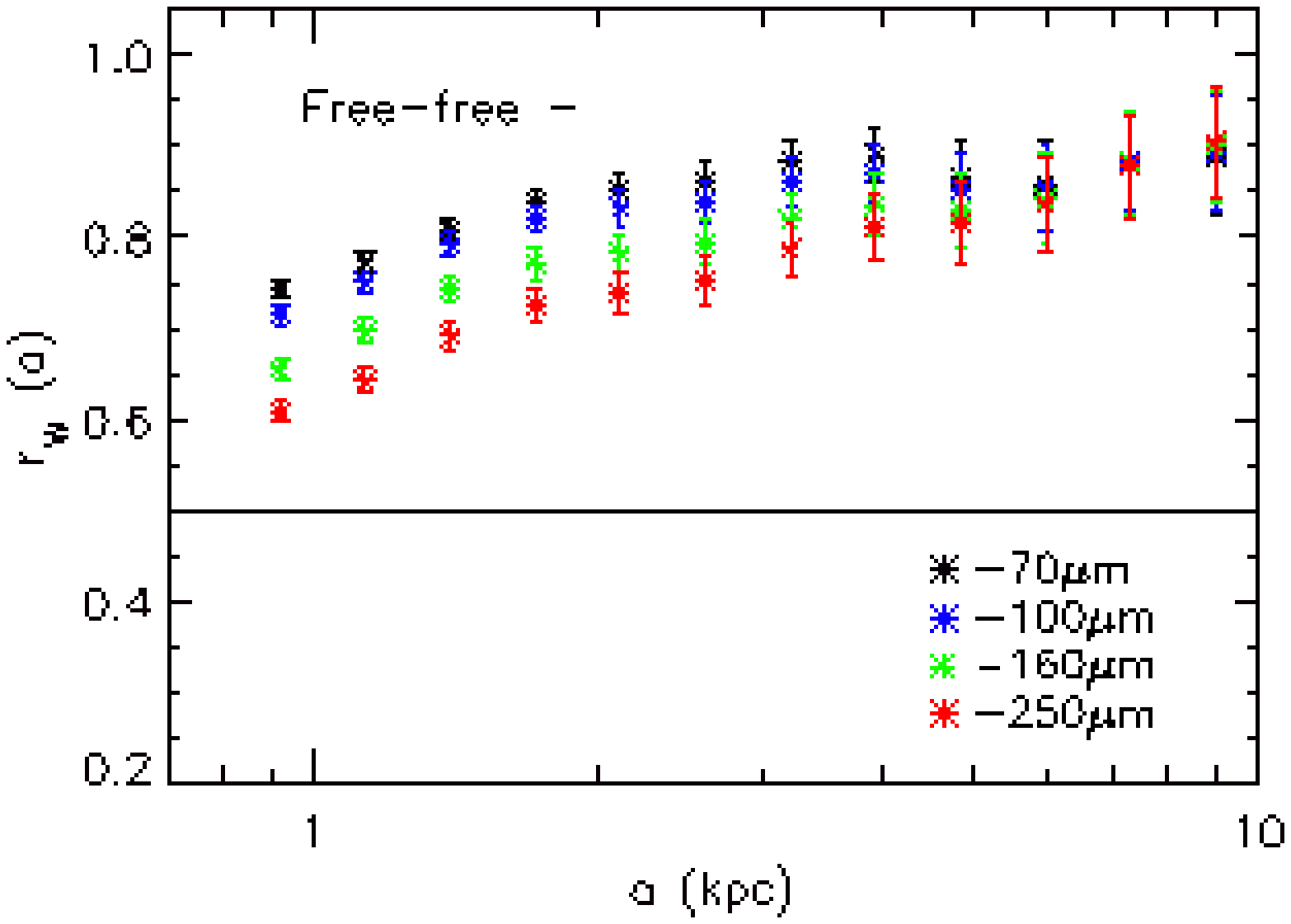}}
\resizebox{\hsize}{!}{\includegraphics*{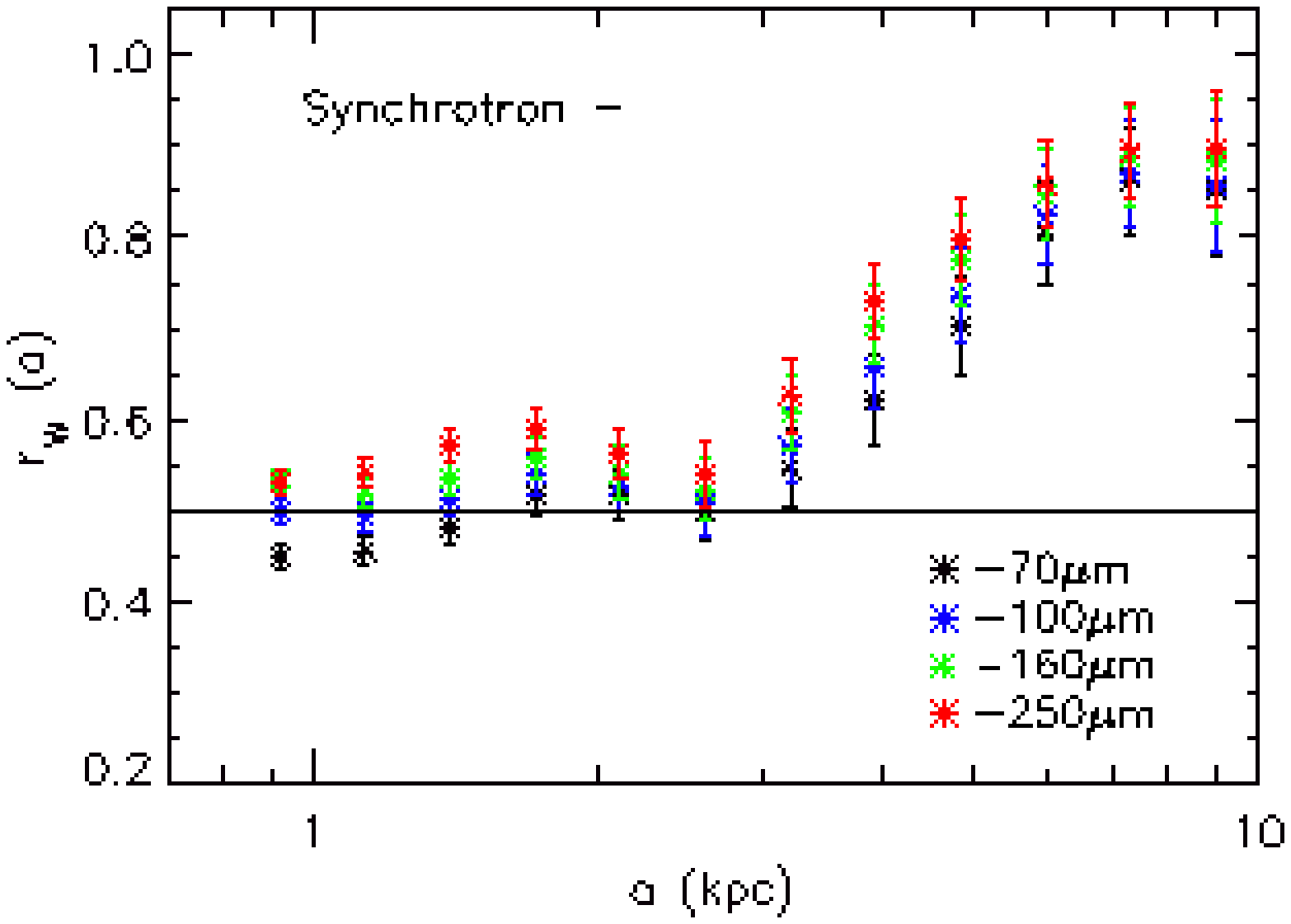}}
\caption[]{The scale-by-scale correlation between the FIR emission and the observed 20\,cm RC (top), the free-free (middle), and the synchrotron (bottom) emissions in NGC\,6946. The horizontal line shows $r_w=0.5$, the threshold for detecting a correlation.  Larger $r_w(a)$ corresponds to more highly correlated structures. These figures illustrate that, on scales smaller than 4\,kpc, the FIR--synchrotron correlation is weaker than the FIR--free-free correlation.}
\end{center}
\end{figure}


\subsection{FIR/radio ratio map}
In order to determine variations in the radio--FIR correlation
across the disk of NGC\,6946, we constructed a logarithmic FIR/radio ratio map using the 20\,cm `synchrotron' map and a FIR$_{42-122\mu{\rm m}}$ \citep[following][]{Helou_etal_85} luminosity map derived by integrating the spectral energy distribution (SED) pixel-by-pixel \citep{Aniano_12}. According to the convention of \cite{Helou_etal_85}, $q$ is defined as: $$q= {\rm log}\,(\frac{\rm FIR}{3.75\,\times\,10^{12}{\rm W}\,{\rm m}^{-2}}) - {\rm log}\,(\frac{S_{1.4}}{{\rm W\,m}^{-2}\,{\rm Hz}^{-1}}),$$ with $S_{1.4}$ the synchrotron 20\,cm flux. The resulting map is shown in Fig.~8. Only pixels detected above the 3$\sigma$ level in each radio and FIR map were considered. The mean $q$ across the galaxy is 2.22$\pm$0.10 with 0.10 the standard deviation. In star-forming regions  we find $q=$\,2.30-2.40 which is in good agreement with the value derived by \cite{Yun_01} for a sample of nearby galaxies (with a mean value of $q=2.34$). In the inter-arm regions, $q$ decreases to values lower than 2.0. Hence, spiral arms
are easily discernible in the $q$ map.
\begin{figure}
\begin{center}
\resizebox{6.8cm}{!}{\includegraphics*{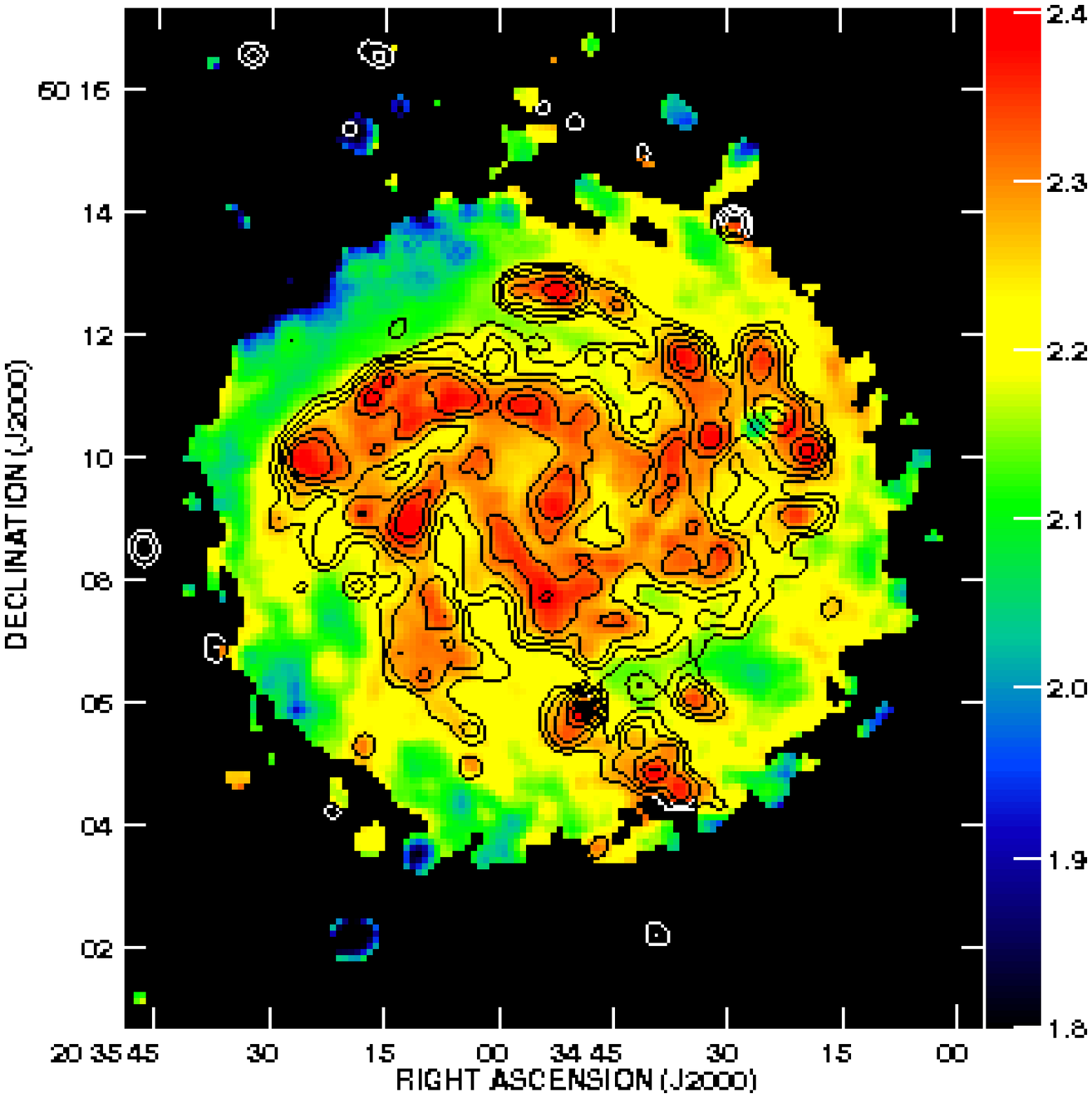}}
\resizebox{6.8cm}{!}{\includegraphics*{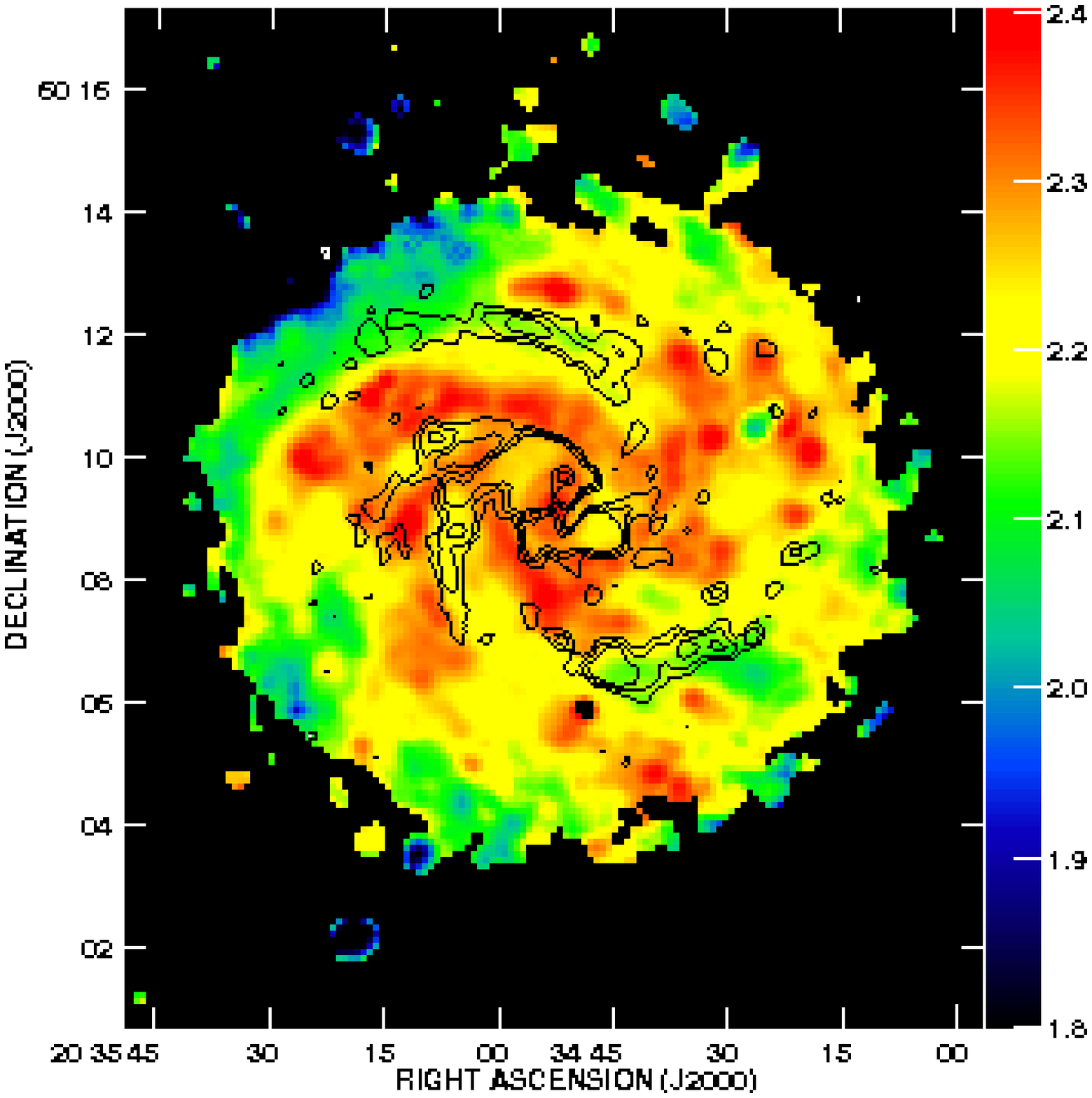}}
\resizebox{6.8cm}{!}{\includegraphics*{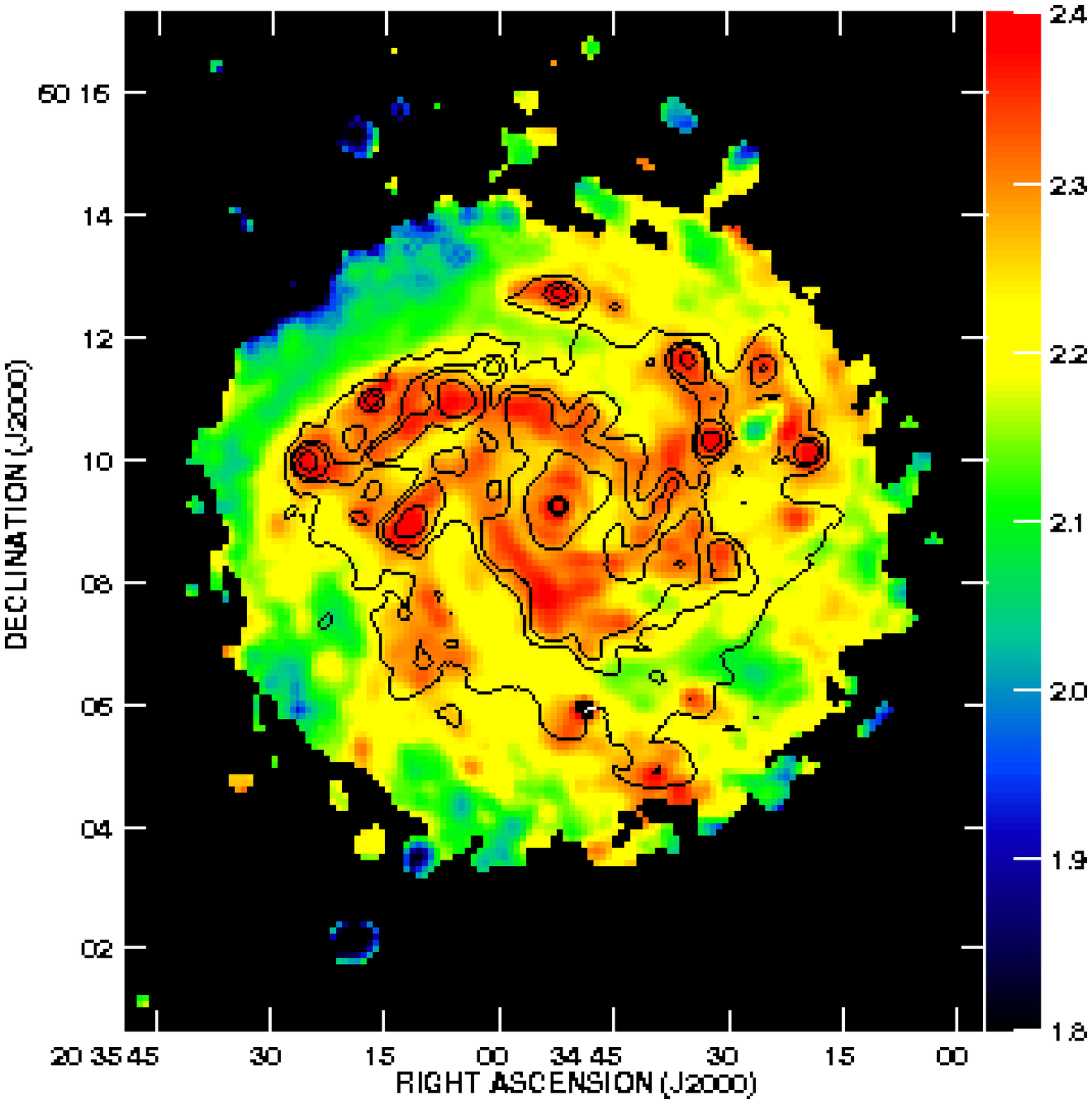}}
\caption[]{The synchrotron-FIR correlation (q) overlaid with contours of the star formation traced by free-free emission with contour levels of 20, 50, 100, 200, 400\,$\mu$Jy/beam ({\it top}), the ordered magnetic field traced by the linearly polarized intensity $PI$ with contour levels of 85, 120, 160\,$\mu$Jy/beam ({\it middle}) and the turbulent magnetic field ($I_{\rm syn} - PI/0.75$) with contour levels of 0.3, 0.6, 1, 2, 25, 32\,mJy/beam ({\it bottom}) in NGC\,6946.  }
\end{center}
\end{figure}

The free-free emission  provides a proper measure for the present-day star formation rate \citep[e.g. ][]{Murphy_11}. The upper panel in Fig.~8 shows a good correlation between $q$ and the free-free emission (contours), in agreement with the important role of SF in controlling the radio--FIR correlation\footnote{We stress that the arm/inter-arm variation of q as well as the dependence of q on SFR is a direct consequence of the nonlinearity of the radio-FIR correlation. Hence, in general, averaging q values could
be misleading.}. 
Besides SF, magnetic fields also seem to play a role as suggested by Fig.~8 (middle) where we observe a correspondence between small $q$ values and peaks of the
polarized intensity.
Furthermore, there is a good agreement between high $q$ values and the turbulent magnetic field (Fig.~8 bottom).  Is there any dependency or competition between SF and magnetic fields in controlling the radio--FIR correlation? A more quantitative comparison between $q$, SFR, and magnetic fields is given in  Sect.~7.

\section{Discussion}
In the following, we discuss the variations in the synchrotron-FIR correlation across the disk of NGC\,6946 as a function of radiation field (important for the dust heating), star formation and magnetic fields. We further investigate possible connections between  SF and the ISM components, magnetic fields and gas. This will help us  to understanding the origin of the observed synchrotron-FIR correlation in this galaxy.   

\subsection{Correlations versus radiation field}
The spatial variations in $q$ may mean that the slope of the correlation is not constant. The higher $q$, e.g., in the arms rather than in the inter-arm regions can be translated as a larger slope ($b$) of the synchrotron-FIR  correlation  in the arms than in the inter-arm regions   \citep[a larger arm/inter-arm slope was also found in M51 by][]{Dumas}.
Similarly, the good correspondence between  $q$ and star formation  (Sect.~7.3) implies that the slope of the synchrotron--FIR correlation is  higher in SF regions and lower in between the arms and outer disk. 
However, if we consider that the synchrotron--FIR correlation originates from massive SF, no
correlation would be expected in regions devoid of SF. Such information on the quality of the correlation cannot be extracted from the $q$ map.
%
In order to investigate both the slope and the quality of the correlation, i.e. the correlation coefficient $r_c$, in SF regions and regions with no SF, we derive the classical correlations in regimes of well-defined properties.

The radiation field is one of the properties which sets the dust heating. Since it is fed by UV photons from the SF regions as well as photons from old and low mass stars (that contribute significantly to the diffuse ISRF), the radiation field could provide a proper defining parameter here.  
Taking advantage of the radiation field map (see Fig.~1, right), we derive the radio--FIR correlation separately for regions where the main heating source of the dust is either SF or the diffuse ISRF. Along the spiral arms and in the central disk, the radiation field has $U > 1.3 U_{\odot}$, with $U_{\odot}$ being the corresponding radiation field in the solar neighborhood,  and it is smaller between the arms and in the outer disk (see Fig.~1). Hence,  $U=1.3 U_{\odot}$ is our  criterion\footnote{This is determined by considering  the radiation field in the solar neighbourhood as a proxy to the diffuse ISRF  optimized for  enough number of points in between the arms and outer disk of NGC\,6946 at 70\,$\mu$m (a 5$\sigma$ level statistically significant correlation or t=11 for 70\,$\mu$m--synchrotron correlation).} to differentiate regimes with ISRF-fed from SF-fed radiation fields. Table~7 lists the results of the  20\,cm synchrotron--FIR correlation analysis for both monochromatic and bolometric FIR emissions in the two defined regimes.
\begin{table*}
\begin{center}
\caption{Synchrotron--FIR/gas correlation in different regimes of radiation fields $U_{\rm ISRF}$ and $U_{\rm SF}$.  }
\begin{tabular}{ l l l l l l l l l } 
\hline
  Y     &  \,\,\,\,\,\, $b^{\rm ISRF}$ &    $r_c^{\rm ISRF}$  & \, t$^{\rm ISRF}$ & \,\,n$^{\rm ISRF}$ &  \,\,\,\,\,\, $b^{\rm SF}$ &    $r_c^{\rm SF}$  & \, t$^{\rm SF}$ & \,\,n$^{\rm SF}$\\
\hline 
\hline
            I$_{70}$ &    0.83$\pm$0.04 & 0.61$\pm$0.05 &  11.0& 209 &  1.38$\pm$0.04 & 0.80$\pm$0.02& 28.4 &451 \\
                  I$_{100}$&  1.00$\pm$0.04   & 0.73$\pm$0.04 & 16.7 & 245  & 1.32$\pm$0.04 & 0.83$\pm$0.03 & 31.7 & 449\\
                 I$_{160}$&    0.90$\pm$0.02 &   0.79$\pm$0.02 & 27.4& 444 & 1.15$\pm$0.04 & 0.85$\pm$0.02 &35.7& 457\\
                  I$_{250}$ &  0.78$\pm$0.02 &  0.80$\pm$0.03 & 23.4 & 396 & 1.07$\pm$0.03 & 0.85$\pm$0.02 & 34.5& 450\\
\hline
                  FIR & 1.05$\pm$0.04 & 0.82$\pm$0.02 & 21.0 & 231 & 1.33$\pm$0.04 & 0.87$\pm$0.02 & 38.6 & 471\\
              $\Sigma_{\rm Gas}$ &  0.68$\pm$0.02 &  0.59$\pm$0.03 & 18.6 & 648 & 1.00$\pm$0.03 & 0.86$\pm$0.02 & 36.1 & 481\\
\hline
$\Sigma_{\rm SFR}$-$\Sigma_{\rm Gas}$&   1.45$\pm$0.06 & 0.58$\pm$0.04 & 14.69 & 320 & 1.34$\pm$0.05 & 0.64$\pm$0.04 & 17.8 & 452\\
B$_{\rm tot}$-$\Sigma_{\rm Gas}$&   0.27$\pm$0.05 & 0.48$\pm$0.05 & 9.7 & 314 & 0.23$\pm$0.01 & 0.75$\pm$0.03 & 23.6 & 429\\

\hline
\end{tabular}
\tablefoot{Pearson correlation coefficients $r_c$ and bisector slopes $b$, log(Y)$\propto b$\,log(X), between the 20\,cm synchrotron (X) and various FIR bands, bolometric FIR, and total neutral gas (Y). The number of independent points, $n$, and  the Student's t-test, $t$, are also indicated for each correlation. Also shown are the correlations between the gas surface density $\Sigma_{\rm Gas}$ and the star formation surface density $\Sigma_{\rm SFR}$, the total magnetic field B$_{\rm tot}$ in logarithmic space. The nucleus was excluded before cross-correlating. }
\end{center}
\label{table:U}
\end{table*}

The slope $b$ of the correlations is indeed shallower in the ISRF than that in the SF regime, in agreement with the $q$ map.  However, apart from the synchrotron correlation with the warm dust (70\,$\mu$m) emission, the correlations in the ISRF regime are as good as those in the SF regime (compare $r_c$ values). This is also the case for the bolometric FIR--synchrotron correlation. 

It is worth mentioning that, among the FIR bands, the 100\,$\mu$m emission has about the same slope as the bolometric FIR versus the synchrotron emission. This can be explained by the fact that the peak of the FIR SED occurs around 100\,$\mu$m in this galaxy \citep{Dale_12}.  

\subsection{Correlations versus magnetic fields}
In the inter-arm regions of NGC 6946, the two magnetic
arms are well traced in the 20 cm synchrotron map (Fig.~3) and show lower $q$ values.
A lower $q$ or a smaller FIR/synchrotron ratio indicates that the synchrotron emission is less influenced  than the FIR emission by the absence of SF in the inter-arm regions. This can be explained by different dependencies of the FIR and synchrotron emission on  SF, or that other mechanisms than SF regulate the synchrotron emission.  The synchrotron emission is a function of the total magnetic field strength and CRE density. Thus, using the synchrotron emission as a tracer of SF implies that both the magnetic field and the CRE density are related to SF. This seems plausible since supernovae produce CREs and induce a turbulent magnetic field via strong shocks in SF regions \citep[e.g. ][]{Reynolds_12}.  However, CREs diffuse and propagate away from their birth places into the magnetized ISM on large scales, where the total magnetic field is dominated by the uniform or ordered field. The origin of the ordered magnetic field can be linked to the dynamo effect on galactic scales \citep[e.g. ][]{beck_90,Beck_96} and is not correlated with  SF\footnote{The ordered magnetic field can also be produced by compressions and shear of the anisotropic or turbulent field in dense gas, known to be the origin of the strong ordered field in the central 6\,kpc gas concentration of NGC\,6946 \citep{Beck_07}} 
\citep[e.g. ][]{chyzy,krause,Fletcher_11}.
Therefore in NGC6946, magnetic arms seem to compensate for the lack of synchrotron emission in the inter-arms as shown by the anti-correlation between $q$ and the linearly polarized intensity PI contours in Fig.~8 (middle panel).
 
%

It is also instructive to investigate the role of the magnetic fields responsible for the synchrotron emission in places with high $q$ (and high SF). Figures 6 and 8 show a good correspondence between high $q$ regions and ${\rm B}_{\rm tur}$, ${\rm B}_{\rm tot}$. 
The scatter plot in Fig.~9 (a) shows that $q$ versus  ${\rm B}_{\rm tot}$ and ${\rm B}_{\rm tur}$ obeys the following power-law relations:
\begin{equation}
q \propto  (0.96 \pm 0.02)\,{\rm log}\,{\rm B}_{\rm tot},  
\end{equation} 
and
\begin{equation}
q \propto (0.85 \pm 0.02)\,{\rm log}\,{\rm B}_{\rm tur}, 
\end{equation}
obtained using the bisector method. Both fits have a dispersion of 0.07\,dex. Hence, the FIR-to-synchrotron flux ratio linearly changes with the total magnetic field strength. Studying similar linearity and also the flatter $q$-${\rm B}_{\rm tur}$ correlation in other galaxies would be of high interest. 

%
\subsection{Correlations versus star formation}
We now aim to characterize the relation between q and SF. We first derive the SF surface density ($\Sigma_{\rm SFR}$) using the de-reddened H$\alpha$ map. The diffuse emission is excluded by masking regions where the free-free 20\,cm flux falls below a threshold value \citep[0.05\,mJy, that is the minimum flux of the detected H{\sc ii} regions by ][]{Lacey}. The nucleus was also subtracted as the optically thin condition does not hold there ($\tau_{{\rm H}\alpha}>1$) and also because of its anomalous properties \citep[for details see ][]{Murphy_11}.  
The de-reddened H$\alpha$ luminosity is converted to SFR using the relation calibrated by
\cite{Murphy_11} for NGC\,6946:
\begin{equation}
\frac{{\rm SFR}}{({\rm M}_{\odot} {\rm yr}^{-1})}=\, 5.37 \times 10^{-42} (\frac{L_{\rm H_{\alpha}}}{ {\rm erg\,s^{-1}}})  \ ,
\end{equation} 
where $L_{\rm H_{\alpha}}$  is the ${\rm H_{\alpha}}$ luminosity density. $\Sigma_{\rm SFR}$ (in units of ${\rm M}_{\odot} {\rm yr}^{-1}\,{\rm kpc}^{-2}$) is then calculated. The power-law behaviour of $q$ versus $\Sigma_{\rm SFR}$ is given by:  
\begin{equation}
q  \propto (0.115 \pm 0.003)\,{\rm log} \,\Sigma_{\rm SFR},   
\end{equation} 
with a Pearson correlation coefficient of $r_c$=0.8  and a dispersion of 0.05\,dex (Fig.~9, b).  Hence, the FIR-to-synchrotron flux ratio changes with a much flatter power-law index with $\Sigma_{\rm SFR}$ than with ${\rm B}_{\rm tot}$  or ${\rm B}_{\rm tur}$.   

\begin{figure*}
\begin{center}
\resizebox{\hsize}{!}{\includegraphics*{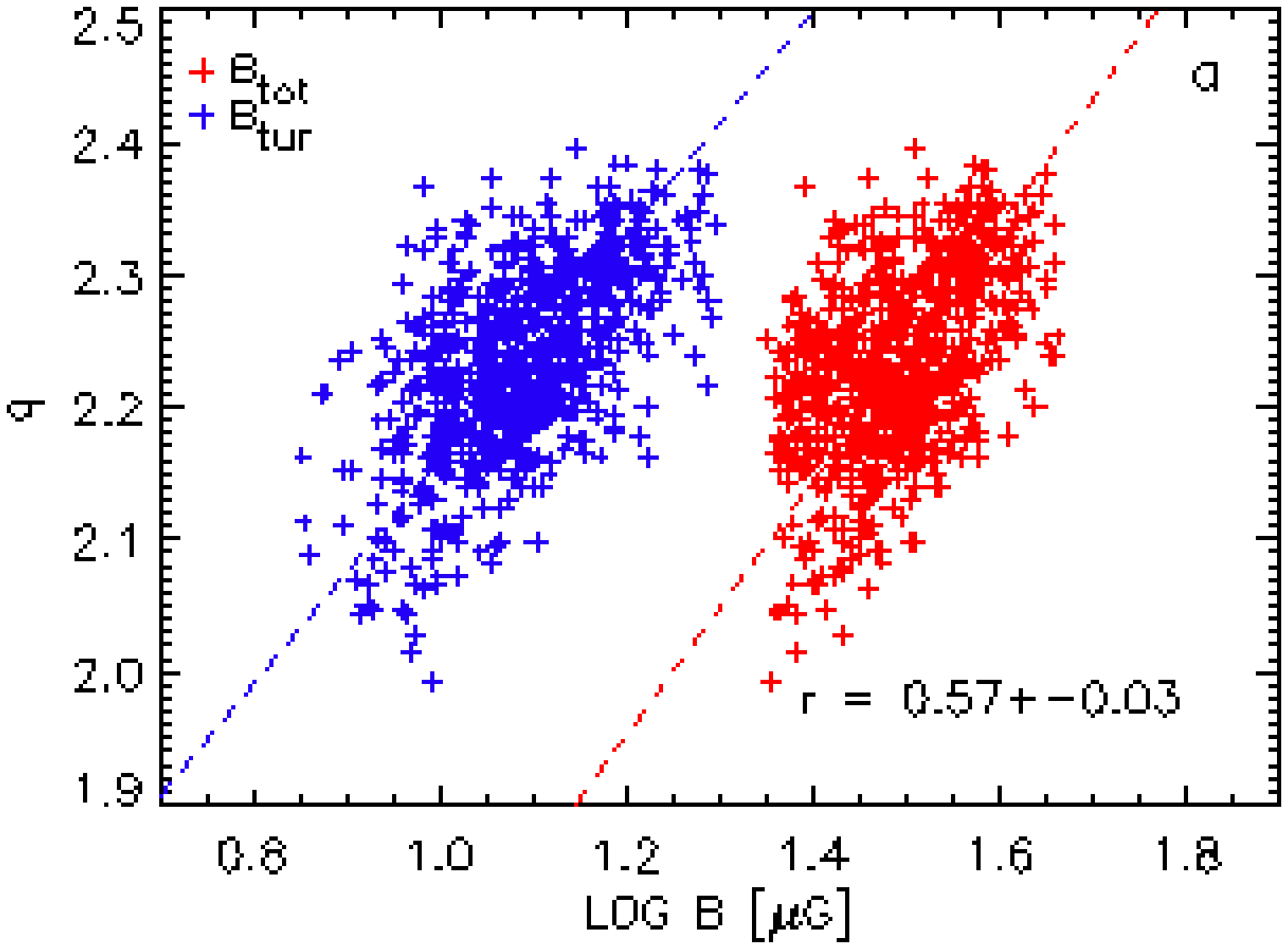}
\includegraphics*{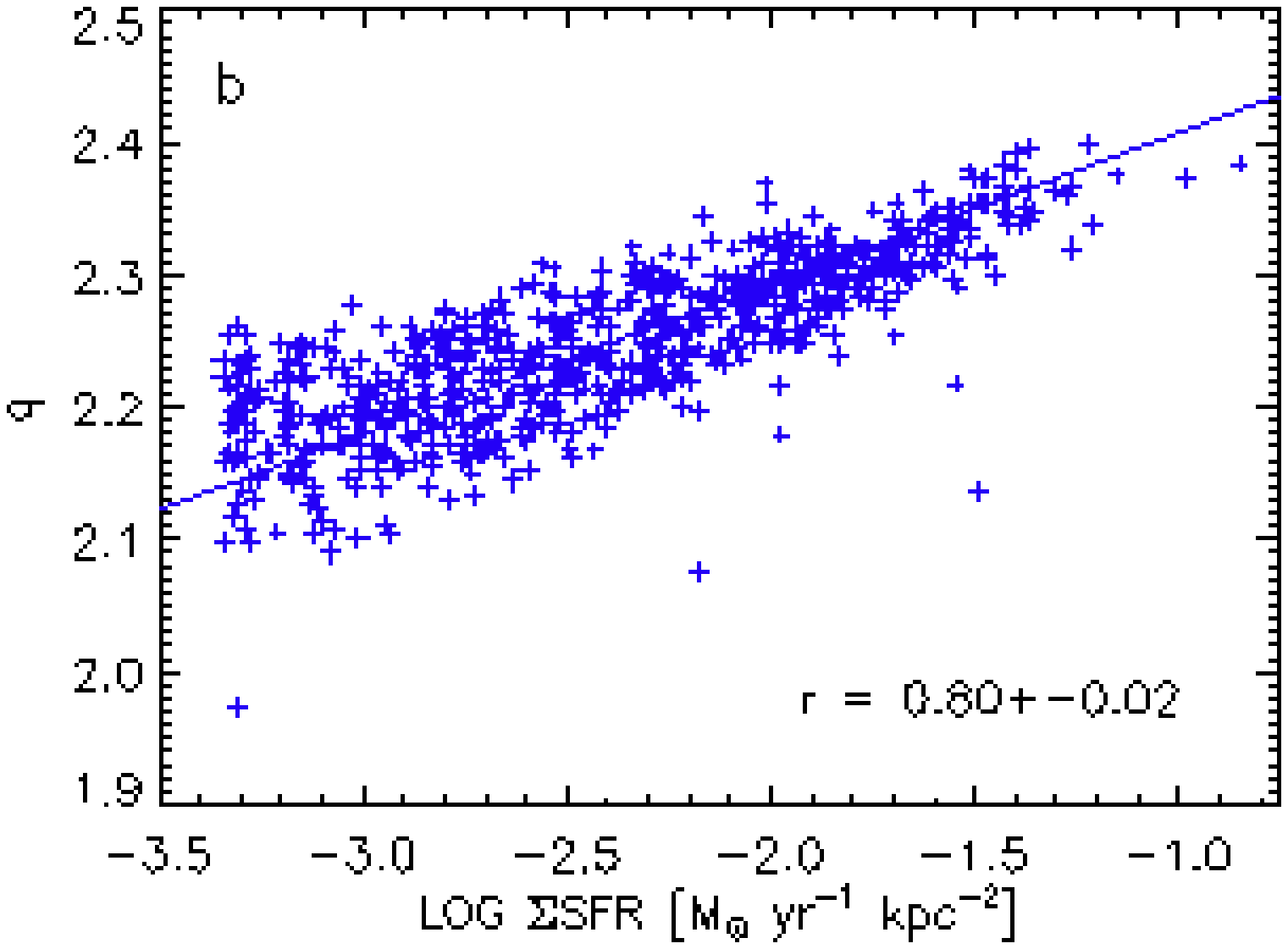}}
\resizebox{\hsize}{!}{\includegraphics*{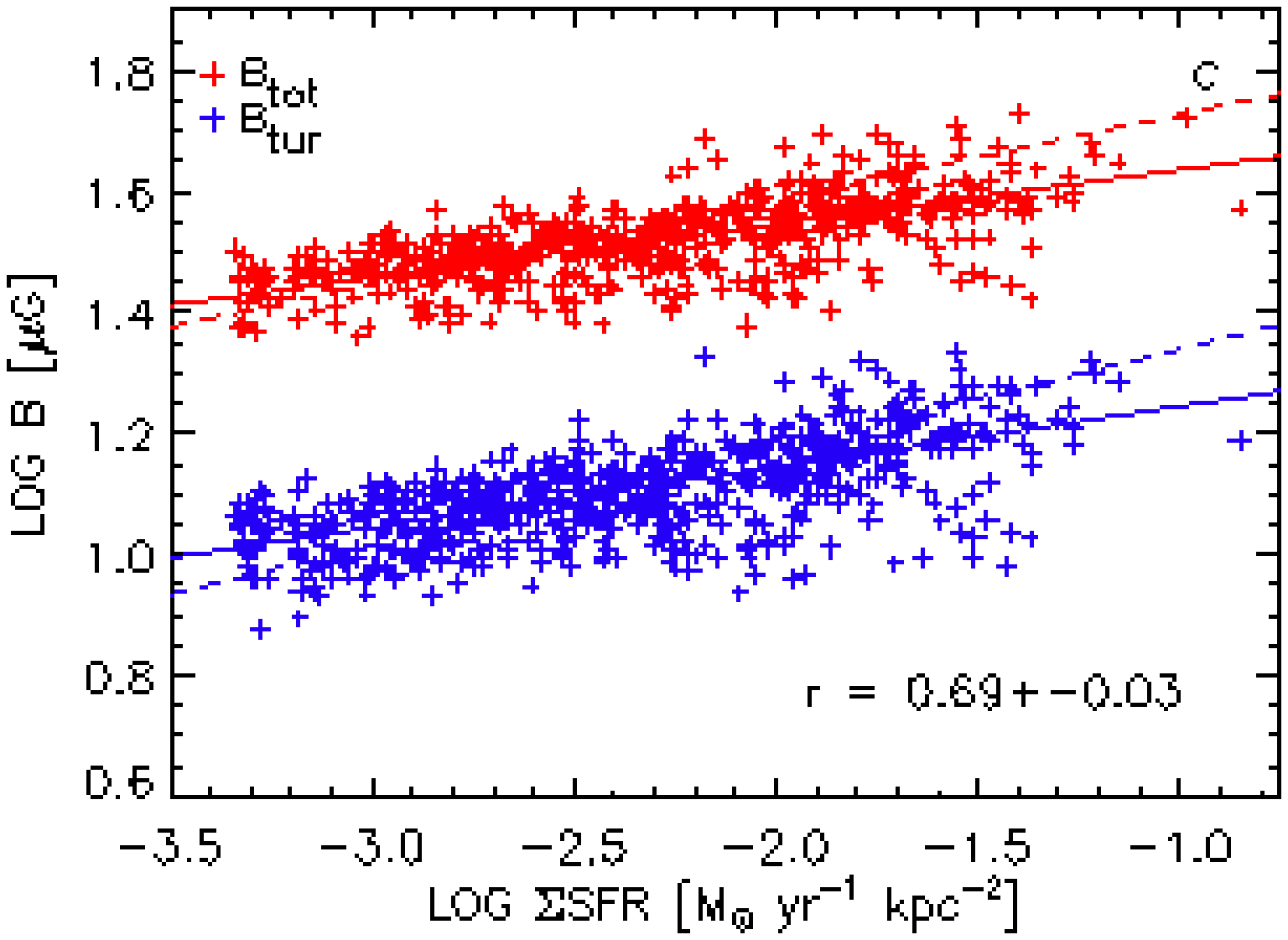}\includegraphics*{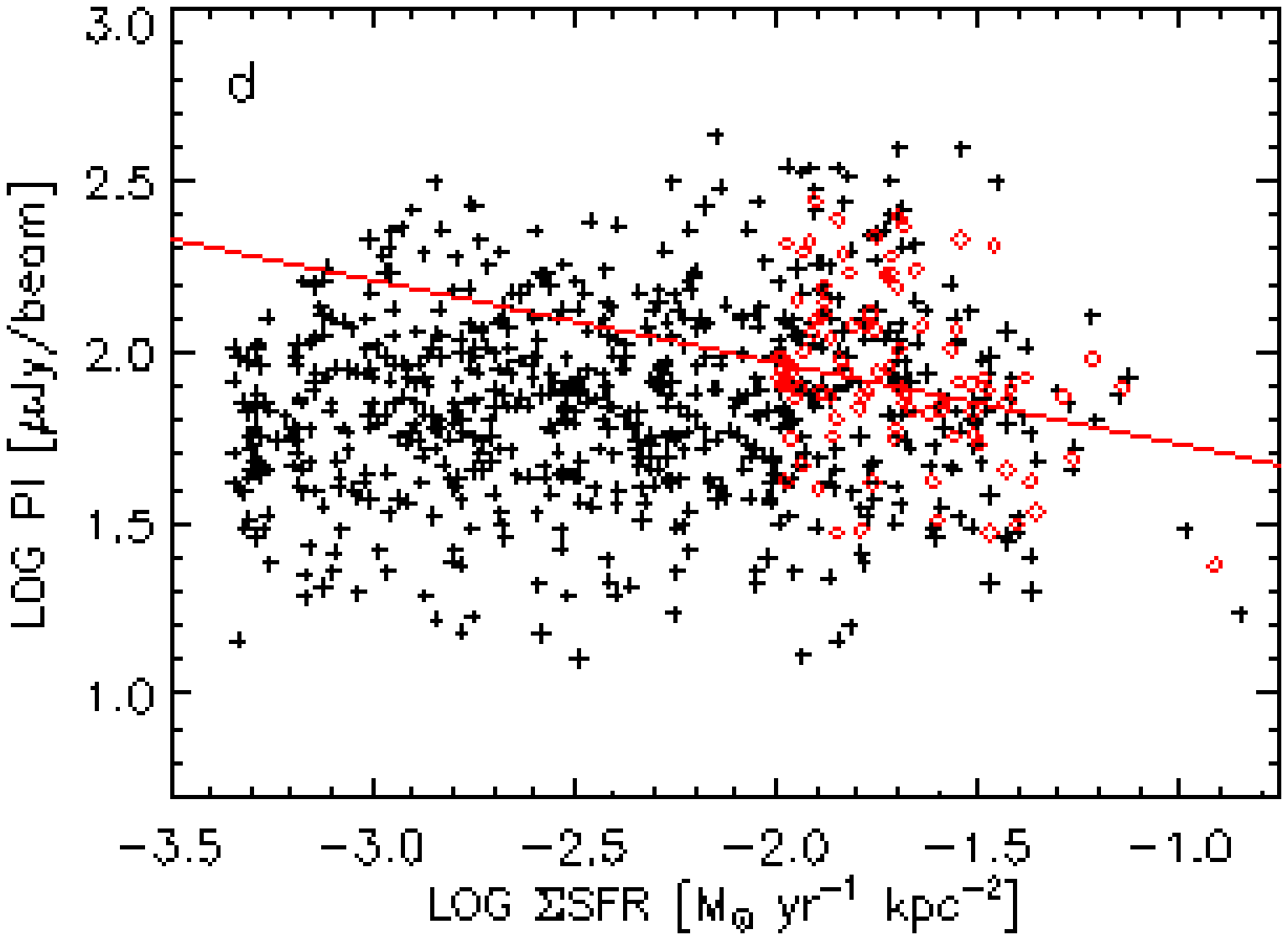}}
\caption[]{The radio-IR correlation (q-parameter) versus total and turbulent magnetic field strength ({\it a}) and star formation surface density $\Sigma_{\rm SFR}$ ({\it b}). The magnetic field strength--$\Sigma_{\rm SFR}$ relation is shown for the total, turbulent ({\it c}), and  the ordered magnetic field strength measured by the degree of polarization ({\it d}). The lines in the upper panels are the bisector fits. The solid lines in {\it c} and {\it d}  show the ordinary least square fit, and the dashed line the bisector fit. In panel {\it d}, the binned data points (diamonds) are fitted for $\Sigma_{\rm SFR}\gtrsim 0.01$\,M$_{\odot}$\,yr$^{-1}$\,kpc$^{-2}$ (solid line). In panels {\it a} and {\it c}, the total magnetic field is shifted up by 0.3 units for clarity. The nucleus is not included in these analysis. }
\end{center}
\end{figure*}

%


\subsection{Connection between star formation and magnetic fields}
As discussed above, star formation can regulate the magnetic field in galaxies.  Therefore, understanding their relationship is important. 
In Fig.~9 (c), we show a good correlation  between $\Sigma_{\rm SFR}$ and both ${\rm B}_{\rm tot}$ ($r_c=0.69\pm0.03$) and ${\rm B}_{\rm tur}$ ($r_c=0.67\pm0.03$).  The bisector fit leads to a power-law proportionality for both ${\rm B}_{\rm tot}$ and ${\rm B}_{\rm tur}$ vs.  $\Sigma_{\rm SFR}$:
\begin{equation}
{\rm log}\,{\rm B}_{\rm tot} \propto  ( 0.14\pm 0.01)\,{\rm log} \, \Sigma_{\rm SFR},   
\end{equation} 
\begin{equation}
{\rm log}\,{\rm B}_{\rm tur} \propto ( 0.16\pm 0.01)\,{\rm log} \,\Sigma_{\rm SFR},   
\end{equation} 
with dispersions of 0.08 and 0.06\,dex, respectively.  Eq.~14 reflects a process where the production of the total magnetic field scales with SF activity. This process probably has a common origin with the production of ${\rm B}_{\rm tur}$ as indicated by the similar slopes in Eqs.~14 and 15. Feedback mechanisms associated with star formation, such as supernova and strong shocks,  produce an increase in turbulence. These could amplify small-scale magnetic fields by a turbulent dynamo mechanism where kinetic energy converts to magnetic energy \citep[e.g.][]{Beck_96,Gressel}. 

On the other hand, there is no correlation between ${\rm B}_{\rm ord}$ (as traced by polarized intensity) and $\Sigma_{\rm SFR}$ for the whole range of parameters ($r_c=0.14\pm0.02$, see the crosses in Fig.~9d). 
However, a weak anti-correlation ($r_c=-0.25\pm0.5$) with a slope of $b=-0.24\pm0.1$ is found for high $\Sigma_{\rm SFR}$ ($\gtrsim 0.01$\,M$_{\odot}$\,yr$^{-1}$\,kpc$^{-2}$), after binning the data points with a width of 0.5  (see the red diamonds in Fig.~9d). 
An anti-correlation for large $\Sigma_{\rm SFR}$ is also visible when comparing the PI and the SF contours in Fig.~8, where  no polarized emission  is associated with the optical spiral arms.  Using the wavelet analysis, an anti-correlation between PI and H$\alpha$ emission was already found by \cite{Frick_etal_01}  on scales equivalent to the width of the spiral arms.
All taken together, this suggests that the production 
of ${\rm B}_{\rm ord}$ could be suppressed along the spiral arms and in SF regions.
An efficient dynamo action in the inter-arm regions could produce a regular field that is anti-proportional to $\Sigma_{\rm SFR}$, as shown for NGC\,6946  \citep[see ][]{Beck_07,Rohde}.

Comparison with a similar study in  another spiral galaxy, NGC\,4254, is instructive. This galaxy is a member of the Virgo galaxy cluster that is  experiencing a gravitational
encounter at the cluster's periphery \citep[e.g.][]{chyzy_07} or  ram pressure due to the galaxy motion through the intracluster medium \citep{Murphy_etal_09} or both of them \citep{Vollmer_05}. Both effects, tidal forces and ram pressures can compress or shear the magnetic fields.
Interestingly,  NGC\,4254 also shows two different trends of ${\rm B}_{\rm ord}$  vs. $\Sigma_{\rm SFR}$, with no general correlation between them \citep{chyzy}. The  division of the two-way behavior, however, occurs at a larger $\Sigma_{\rm SFR}$ ($\simeq 0.02$\,M$_{\odot}$\,yr$^{-1}$\,kpc$^{-2}$) than in NGC\,6946. 
Moreover, in NGC\,4254, the slope of ${\rm B}_{\rm tur}$-$\Sigma_{\rm SFR}$ relation (0.26$\pm$0.01) is larger than  that of ${\rm B}_{\rm tot}$-$\Sigma_{\rm SFR}$ relation (0.18$\pm$0.01), unlike in NGC\,6946.  The slope of ${\rm B}_{\rm tur}$-$\Sigma_{\rm SFR}$ relation is also larger in NGC\,4254 than in NGC\,6946 (0.16$\pm$0.01, see Eq.~15). 
Hence, although the two galaxies resemble each other in their general magnetic field--$\Sigma_{\rm SFR}$ behavior,  they differ in the details of these connections. 
%
These differences could be related to  different contributions of ${\rm B}_{\rm ord}$  and ${\rm B}_{\rm tur}$ to ${\rm B}_{\rm tot}$, as a similar scaling relation holds between ${\rm B}_{\rm tot}$ and $\Sigma_{\rm SFR}$ in these galaxies.  Hence, investingating ${\rm B}_{\rm ord}$-to-${\rm B}_{\rm tur}$ ratio (${\rm B}_{\rm ord}/{\rm B}_{\rm tur}$) could be instructive. In NGC\,4254,  ${\rm B}_{\rm ord}/{\rm B}_{\rm tur}$ shows a general decreasing trend with increasing $\Sigma_{\rm SFR}$ for the whole range of values \citep[see Fig.~7b of][]{chyzy}. In NGC\,6946, such  a decreasing trend  is indicated only for $\Sigma_{\rm SFR} \gtrsim 0.01$\,M$_{\odot}$\,yr$^{-1}$\,kpc$^{-2}$ (Fig.~10).  For small $\Sigma_{\rm SFR}$ values, ${\rm B}_{\rm ord}/{\rm B}_{\rm tur}$ is larger in NGC\,4254 than in NGC\,6946. 

In the Virgo Cluster galaxy NGC\,4254, the strongest ordered field is found in the outer arms, dominated by anisotropic turbulent field and is probably a product
of shearing/stretching forces caused by weak gravitational interaction \citep{chyzy} and/or ram pressure \citep{Murphy_etal_09} in the cluster environment. Such forces 
can transform ${\rm B}_{\rm tur}$ into ${\rm B}_{\rm ord}$ in the outer arms of NGC\,4254 where $\Sigma_{\rm SFR}$ is low (particularly in the southern arm). This could cause the steeper log(${\rm B}_{\rm tur}$)-log($\Sigma_{\rm SFR}$) than  log(${\rm B}_{\rm tot}$)-log($\Sigma_{\rm SFR}$) relation in NGC\,4254.

In spite of the magnetic arms, NGC6946 yields smaller values
of ${\rm B}_{\rm ord}/{\rm B}_{\rm tur}$ at low SFR compared to NGC4254 
because it does not experience shearing due to tidal forces and/or ram pressures inserted from a cluster environment. The total field has similar properties in both galaxies (same slope), because the shearing does not change the total field strength. It only transforms
turbulent into ordered field. This could explain the flatter slope of log(${\rm B}_{\rm tur}$)-log($\Sigma_{\rm SFR}$) in NGC6946 than that in NGC4254. 

Therefore, we might expect that, in normal galaxies like NGC6946, the slope of log(B)  vs. log($\Sigma_{\rm SFR}$) is similar for both ${\rm B}_{\rm tur}$ and ${\rm B}_{\rm tot}$, while it is steeper for  ${\rm B}_{\rm tur}$ than for  ${\rm B}_{\rm tot}$ in galaxies interacting with the cluster environment (like NGC4254). 
\begin{figure}
\begin{center}
\resizebox{7cm}{!}{\includegraphics*{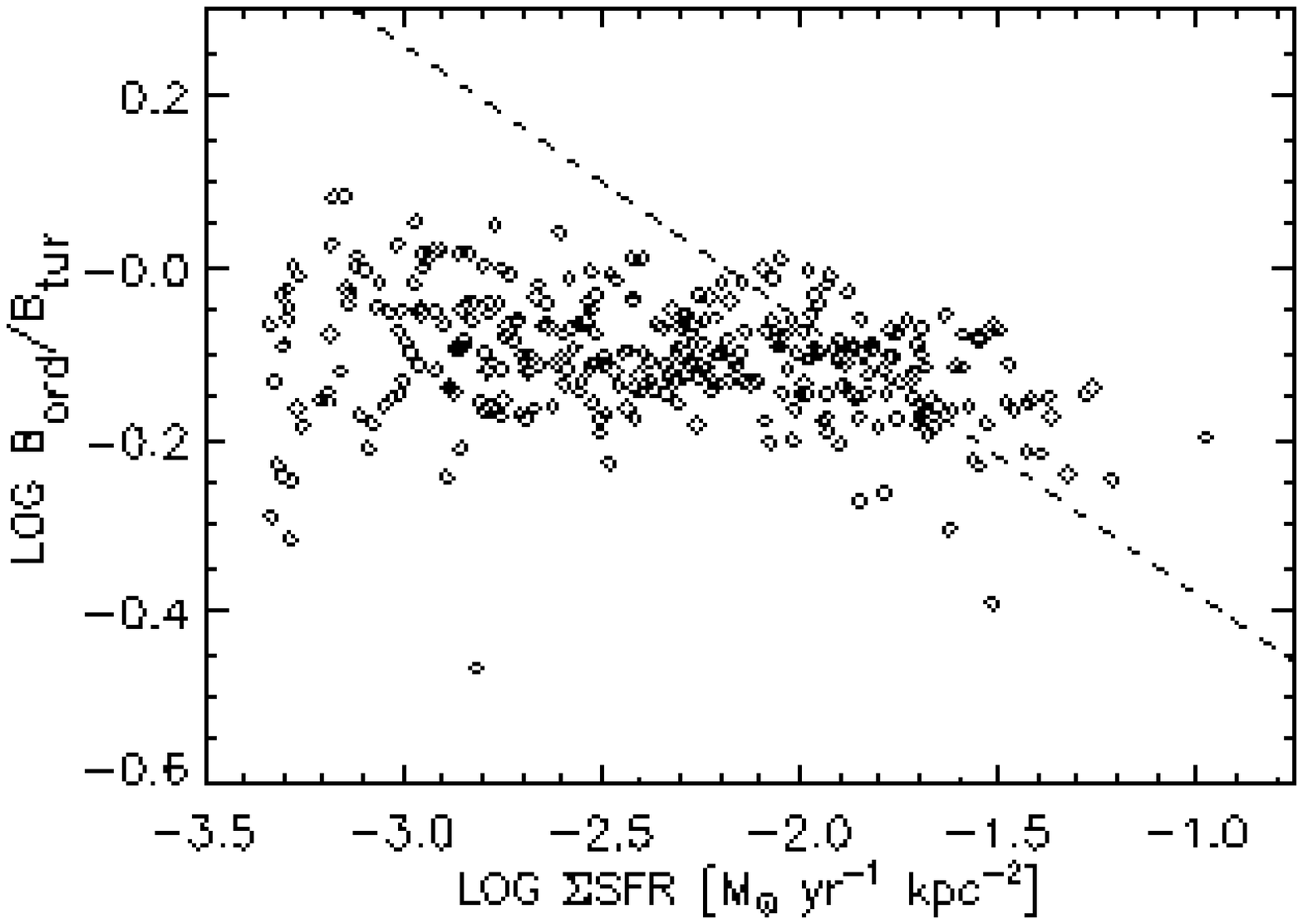}}
\caption[]{ The ratio of the Ordered-to-turbulent magnetic field strengths B$_{\rm ord}$/B$_{\rm tur}$ vs. star formation surface density $\Sigma_{\rm SFR}$ in NGC\,6946. The dashed line shows the corresponding fitted line in NGC\,4254 shifted by 0.2 along the x-axis  \citep[$y=-0.32\,x-0.7$, see Fig.~7b in][]{chyzy}. }
\end{center}
\end{figure}

\begin{figure}
\begin{center}
\resizebox{7cm}{!}{\includegraphics*{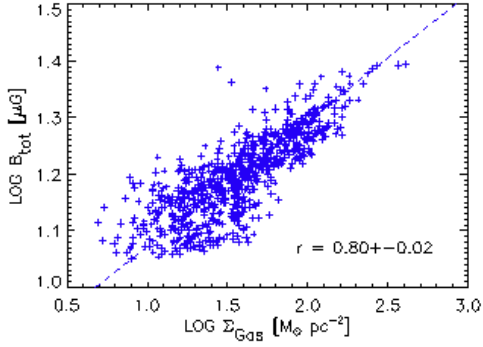}}
\caption[]{ Correlation between the magnetic field and the total gas surface density $\Sigma_{\rm Gas}$. Also shown is the bisector fit. The nucleus is not included in this analysis. }
\end{center}
\end{figure}
\subsection{Correlations with neutral gas }
Some of the radio--FIR correlation models are based on a coupling between the magnetic field B and gas density $\rho$ \citep{Helou_93,Niklas_97,Hoernes_etal_98,Murgia}.     
Physically, the B-$\rho$ coupling could be caused by the amplification of the Magneto Hydrodynamic (MHD) turbulence until energy equipartition is reached \citep{groves_03}. 
This model has the advantage of 1) producing the radio--FIR correlation where the FIR is dominated by dust heated by older stellar populations  or the ISRF \citep[e.g. ][]{Hoernes_etal_98} and 2)  naturally producing a $local$ radio--FIR correlation, breaking down only on scales of the diffusion length of the CREs $l_{\rm dif}$ \citep[as discussed in ][]{Niklas_97}.  
 
Table~7 shows that the total molecular and atomic gas surface density $\Sigma_{\rm Gas}$ in  M$_{\odot}$\,pc$^{-2}$ is correlated with the synchrotron emission in both the SF and the ISRF regimes. The correlation, in the ISRF regime, is however not as tight as in the SF regime (see $r_c$ values in Table~7)  because the inter-arm regions (heated by the ISRF) are filled with low-density atomic HI gas which appears to have a weaker correlation with synchrotron emission compared to dense molecular gas \citep[][]{Beck_07}. In M51,  \cite{Tilanus} found a much weaker correlation between the diffuse HI gas and the synchrotron emission.  

A  tight correlation between $\Sigma_{\rm Gas}$ and the total magnetic field holds for the entire galaxy (Fig.~11). The total magnetic field B$_{\rm tot}$ vs. $\Sigma_{\rm Gas}$ follows a power-law relation, 
\begin{equation}
{\rm log}\,{\rm B}_{\rm tot} \propto ( 0.23\pm 0.01)\,{\rm log} \, \Sigma_{\rm Gas}.
\end{equation} 
The dispersion of the fit is about 0.04. In the ISRF regime, the B$_{\rm tot}$--$\Sigma_{\rm Gas}$ correlation is weaker than in the SF regime, although the power-law indices are similar (Table~7).                          

\cite{Niklas_97} suggested that the tight radio--FIR correlation can be reached by considering the coupling between the FIR emission and the gas density via the  Kennicutt-Schmidt (KS) relation between $\Sigma_{\rm SFR}$ and $\Sigma_{\rm Gas}$ \citep{Schmidt,Kennicutt_98}. Interestingly, in NGC\,6946, not only  does the KS  relation  hold, but also it is very similar in the ISRF and  SF regimes (Table~7 and Fig.~12). Hence, it seems that in the ISRF regime, the lower gas density and the weaker radiation field conspire to hold the KS  relation but with a weaker $\Sigma_{\rm SFR}$--$\Sigma_{\rm Gas}$ correlation.
%
In the SF regime, the  $\Sigma_{\rm SFR}$--$\Sigma_{\rm Gas}$ power-law index of $\sim$\,1.3 is in agreement with \cite{Kennicutt_98}. The KS index in both regimes agrees with the global value of 1.46$\pm$0.29 for NGC\,6946 presented in \cite{Bigiel_08}.

An equipartition between the CRE and magnetic field energy densities is needed in the B-$\rho$ model of \cite{Niklas_97}.  This implies that the energy density of CREs (and the synchrotron luminosity)  is determined by the field strength alone, and neither the CRE production rate nor the CRE escape probability affects it. On small spatial scales, however, e.g. in supernova remnants, the energy density of the particles may exceed strongly the energy density of the magnetic field.  A static pressure equilibrium could then be achieved when the components of the interstellar medium have been relaxed on larger scales (e.g. determined by a scale height of gas clouds of a few 100 pc). Hence, the local synchrotron--FIR correlation should break down on small spatial scales where equipartition is not valid anymore.
Looking for such a break in the synchrotron--FIR correlation in NCG\,6946, we refer to our scale-by-scale correlation study. Figure~7 shows that the smallest scale on which the synchrotron--70\,$\mu$m correlation holds, i.e. $r_w\sim0.5$,  is $a\sim1.7$\,kpc (on scales smaller,  $r_w<0.5$     and hence it is below the threshold value for the acceptability of the correlation). This scale can be translated as  $l_{\rm dif}$ or the scale of the static pressure equilibrium, according to the B-$\rho$ model \citep{Niklas_97}. However, this is not certain, since the colder dust traced by the longer FIR wavelengths do not show similar break $r_w<0.5$ when correlating with the synchrotron emission. 
Nevertheless, the synchrotron--70\,$\mu$m correlation has been used in previous studies to determine the CRE diffusion length. For instance, using the Spitzer MIPS 70\,$\mu$m data, \cite{Murphy_08} estimated $l_{\rm dif}$ in a sample of nearby galaxies by applying an image-smearing method. Using the same FIR wavelength (70\,$\mu$m), our derived $l_{\rm dif}$  agrees with the best-fitted disk value of \cite{Murphy_08} in NGC\,6946, $l_{\rm dif}=1.6 \pm 0.1$\,kpc,  although the two methods are different.  
\begin{figure*}
\begin{center}
\resizebox{12cm}{!}{\includegraphics*{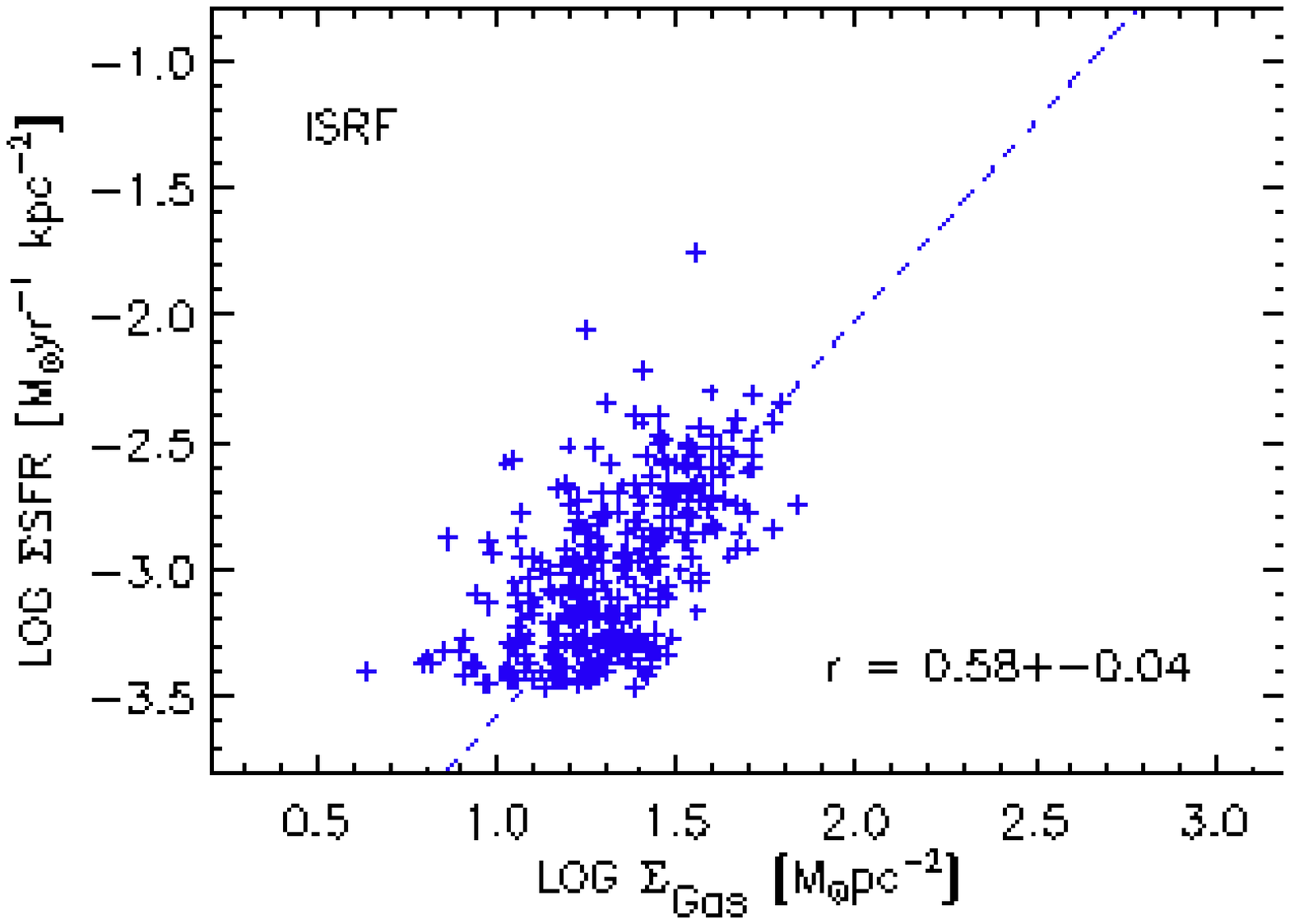}
\includegraphics*{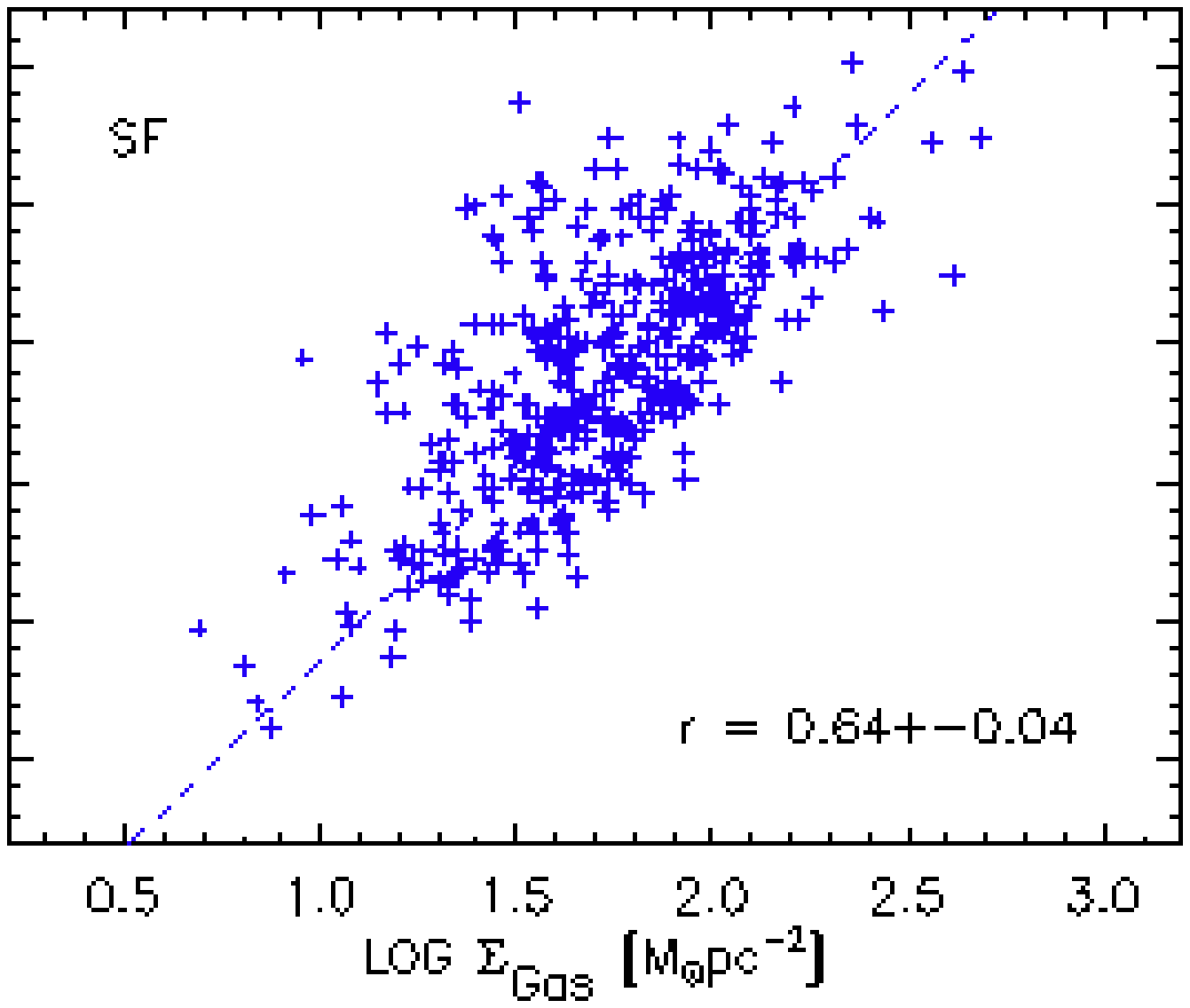}}
\caption[]{Correlations between the $\Sigma_{\rm SFR}$ [M$_{\odot}$\,yr$^{-1}$\,kpc$^{-2}$] and the total gas surface density $\Sigma_{\rm Gas}$ [M$_{\odot}$\,pc$^{-2}$] for the ISRF (left) and SF regimes (right) as defined by $U>1.3\,U_{\odot}$. Also shown are the bisector fits. The nucleus is not included in this analysis.}
\end{center}
\end{figure*}
\subsection{The origin of the radio--FIR correlation in NGC\,6946}
%
 %
The proposed models to explain the radio--FIR correlation in galaxies include: \\
a)  The calorimeter model of \cite{Voelk}. This model assumes that CREs are completely trapped in their host galaxies and that the galaxies are optically thick for the dust-heating stellar UV photons. In the case of complete CRE trapping, a synchrotron spectral index of $\alpha_n\geq$\,1.0 is expected, i.e. synchrotron and inverse Compton losses are  the dominant cooling mechanism of CREs.  \\
%
b) The optically-thin model of \cite{Helou_93} and \cite{Niklas_97}. The dust-heating UV photons and the
cosmic rays are assumed to have a common origin in massive star formation similar to the model of \cite{Voelk}, 
but in this case both photons and cosmic rays can escape the galaxy. Although  different distribution for the CREs energy density is applied by \cite{Helou_93} and \cite{Niklas_97} \citep[also see][]{Hoernes_etal_98,Murgia}, the basic condition of B-$\rho$ coupling is a common assumption in these models, as discussed in Sect.~7.5.  \\
c) The calorimeter model of \cite{Lacki_10} in starbursts, in which cosmic ray protons lose all of their energy and produce secondary electrons and positrons whose synchrotron emission keeps the radio--FIR correlation linear. This was motivated by \cite{Thompson_06} who suggested that Bremsstrahlung and ionization losses are more important in starburst galaxies leading to a flat $\alpha_n$ ($\simeq$0.75).  According to \cite{Lacki_10}, the CRE density is not proportional to the SFR in starbursts, in contrast to both the calorimeter model of \cite{Voelk} and the `optically-thin' models of \cite{Helou_93} and \cite{Niklas_97}, who explained the observed non-linear synchrotron--FIR correlation in the Shapley-Ames galaxies \citep[presented in ][]{Niklas_977} by assuming a SFR related synchrotron emission and B-$\rho$ coupling.  

In NGC\,6946, the complete CRE trapping assumed in the model (a)  does not apply, since depending on the location, the synchrotron spectral index changes between a flatter $\alpha_n \simeq$\,0.6 (in the star-forming regions) and a steeper $\alpha_n \simeq$\,1 (in regions of strong ordered magnetic field, Sect.~4). This indicates the presence of various loss mechanisms of the CREs energy leading to a flatter nonthermal spectrum (mean $\alpha_n \simeq$0.8) than that produced by the efficient synchrotron and inverse Compton losses ($\alpha_n$=\,1.0-1.2) for the entire galaxy. 

Although NGC\,6946 is classified as a normal star-forming galaxy, it would be instructive to see its position in respect to the model (c). The synchrotron--FIR correlation is non-linear with the warmer dust globally (Table~5), or  the dust heated by the SF (Table~7). On the contrary, a linearity is found for the colder dust globally (Table~5), or the dust heated by the ISRF (Table~7). The linearity of the correlation cannot be explained by secondary electrons \& positrons suggested for dense gas condition in starbursts, due to the low gas density of the ISRF regime in NGC\,6946 \citep[in agreement with][]{Murphy_09}. 

As shown in Sect.~7.5, we already obtained indications for a B-$\rho$ coupling in NGC\,6946 which is in favor of the model (b). How this model could also reproduce the observed slope of the synchrotron--FIR correlation in the SF and the ISRF regimes are detailed as follows:
%
\\
In the SF regime, the KS relation gives $\Sigma_{\rm SFR}~\propto~\Sigma_{\rm Gas}^{\rm KS}$, with the KS index of 1.3$\pm$0.05 (Table~7). Our data show that the FIR luminosity is proportional to $\Sigma_{\rm SFR}$ with a slope of 0.95$\pm$0.05. Hence, 
$${\rm FIR} \propto \Sigma_{\rm Gas}^{1.23 \pm 0.07}.$$ 
We note that the same proportionality is derived by directly correlating between the FIR luminosity and $\Sigma_{\rm SFR}$, as expected. From Eq.~(6), the equipartition magnetic field is given by: 
$${\rm B}_{\rm tot} \propto I_{\rm syn}^{1/(3+\alpha_n)},$$ 
with $\alpha_n=0.8\pm0.1$ (Sect.~4). 
In Sect.~8.2, we found that 
$${\rm B}_{\rm tot} \propto \Sigma_{\rm Gas}^{0.23\pm 0.01}.$$ 
Following \cite{Niklas_97} and \cite{Hoernes_etal_98}, we assume that the scale height of the neutral gas is equal to that of the dust and is constant. Thus, $\Sigma_{\rm Gas}$ can be replaced by the gas volume density in the above proportionalities resulting in 
\begin{equation}
{\rm FIR}\,\propto\,I_{\rm syn}^{1.41 \pm 0.12}.
\end{equation}
This is in excellent agreement with the observed synchrotron--FIR correlation in the SF regime with the slope of $b^{\rm SF}=1.33\,\pm$\,0.04 (Table~7).

A similar calculation for the ISRF regime leads to 
\begin{equation}
{\rm FIR}\, \propto\,I_{\rm syn}^{1.34 \pm 0.14}.
\end{equation} 
The slope is higher than the observed slope of $b^{\rm SF}~=~1.05\,\pm$\,0.04 by about 2$\sigma$  errors (28\%). 

\cite{Xu_94} found a linear correlation between the 20\,cm synchrotron and cold dust emission in a sample of  late-type spirals. They explained this correlation by considering intermediate mass stars (5-20\,$M_{\odot}$) as a heating source of cold dust and synchrotron emission, since these stars are supernova progenitor as well. It is also likely that, in NGC\,6946 which is a late-type spiral with a good 20\,cm synchrotron--cold dust correlation  (e.g. see Table~5), the intermediate mass stars provide the non-ionizing UV photons to heat the cold dust (which emit at longer FIR wavelengths). As such, these UV photons provide the bulk of the ISRF in NGC\,6946.   

In the previous sections, we showed that the local radio--FIR correlation varies as a function of not only star formation rate  but also dust heating sources, magnetic fields,  and gas density; i.e., the ISM properties. The SF and ISM are continuously influencing each other: Stars form within the dense and cold regions of the ISM, molecular clouds, and replenish the ISM with matter and energy.   
Hence, {\it the local radio--FIR correlation is a probe for the SF--ISM interplay} which is partly explained through scaling relations such as the KS relation between $\Sigma_{\rm SFR}$ and gas surface density and the relation between $\Sigma_{\rm SFR}$ and the magnetic field strengths (Sect.~7.4).

The {\it global radio-FIR correlation} is known to be a tracer of SF. 
Among the SF-based theories to explain  the global radio-FIR correlation, our local studies match with those  considering a coupling between the magnetic field and gas density, as the radio--FIR correlation also holds  in regions with no massive SF, e.g. in the inter-arm regions and the outer disk (Sect.~7.1). This shows that a balance between the gas and  magnetic field/CRE pressure is an unavoidable condition for the correlation. Apart from this, the global radio--FIR correlation as a tracer of SF still applies since  the integrated radio and FIR fluxes are weighted towards more luminous regions of galaxies, i.e. the SF regions. 
\subsection{ISM and propagation of CREs}
%
Propagating through the ISM, CREs can experience various energy losses via ionization, bremsstrahlung, adiabatic, synchrotron, and inverse-Compton losses that change the power-law index of the energy distribution of  these particles or equivalently the nonthermal spectral index, $\alpha_n$. 
The maps of the synchrotron spectral index (Sect.~4),  magnetic fields (Sect.~5),  and radiation field (Sect.~3), provide    direct information on the main cooling mechanisms of CREs, as well as the cooling timescale and diffusion scalelength  of  CREs in NGC\,6946. 
As shown in Fig.~4, there is a difference in the spectral indices of $\Delta \alpha_n\,\simeq$\,0.5 between the star-forming regions and the magnetic arms/inter-arm regions. This is expected if the main mechanisms of energy losses for the electrons are synchrotron emission and inverse Compton scattering \citep{Longair}.  The CRE cooling timescale $t_{\rm cool}$  (in units of yr) associated with the two dominant processes can be derived from the following formula \citep{Murphy_08}:
%
\begin{eqnarray}
t_{cool} & \sim &   5.7 \times 10^7  \left. \left(\frac{\nu_c}{\rm GHz}\right)^{-0.5} \times \, \right.
\nonumber \\
& &\left.  (\frac{\rm B}{\mu{\rm G}})^{0.5} \, \big(\frac{U_B + U_{\rm rad}}{10^{-12}\, {\rm erg\,cm^{-3}} } \big)^{-1} \right.,
\end{eqnarray}
%
where $\nu_c \sim {\rm B} E^2$ is the critical frequency at which a CRE emits most of its  energy $E$. $U_B={\rm B}^2/(8 \pi)$ is the magnetic field energy density, and $U_{\rm rad}$ is the radiation energy density. 

Using a simple random walk equation, CREs will diffuse over a distance $l_{\rm cool}= (D_E \,t_{\rm cool})^{0.5} $  before losing all of their energy to synchrotron and inverse Compton losses, with the energy-dependent diffusion coefficient $D_E$.
Assuming that the diffusion  length scale of 1.7\,kpc, obtained from the synchrotron--FIR correlation (Sect.~7.5), is equivalent to the cooling scale length $l_{\rm cool}$,  $D_E$ can be estimated independently. \cite{Mathis_83} determined the radiation field energy density in starlight of  $U_{\rm rad}=\,8.64\,\times\,10^{-13}\,U$\,erg\,cm$^{-3}$. However, $U_{\rm rad}$ should also include the dust emission energy density, $U_{\rm FIR}=\,5\,\times\,10^{-13}\,U$\,erg\,cm$^{-3}$ as well as the cosmic microwave background radiation energy density, $U_{\rm CMB}=\,4.17\,\times\,10^{-13}$\,erg\,cm$^{-3}$ \citep[][]{Draine_11}.
Therefore, we use a more general form of $U_{\rm rad}=\,(4.17\,+\,13.64\,U) \times\,10^{-13}$\,erg\,cm$^{-3}$. 
For the entire galaxy, $U \simeq$2\,$U_{\odot}$ and B\,$\simeq 18$\,$\mu$G. Thus, $U_B\simeq 1.3 \times 10^{-11}$\,erg\,cm$^{-3}$ and $U_{\rm rad} \simeq 3.1 \times 10^{-12}$ erg\,cm$^{-3}$.    
Substituting these values in Eq.~(19) results in $t_{\rm cool} \simeq 1.3\times 10^{7}$ yr. The CRE diffusion coefficient is then $D_E\simeq 6.8\times 10^{28}$\,cm$^2$\,s$^{-1}$. Assuming that $D_E$ changes with energy as $D_E=\,D_0\,(E/GeV)^{0.5}$ \citep[e.g.][]{Ginzburg}, we derive the normalization factor $D_0=4.6\times 10^{28}$\,cm$^2$\,s$^{-1}$ for the 2.2\,GeV CREs. 

As discussed in Sect.~6.2, the drop in the synchrotron--FIR correlation on  the larger scales of  about 2.5\,kpc is related to a lack of FIR emission from the magnetic arms (in the inter-arm regions), where diffused CREs experience synchrotron loss. Hence, we assume that the cooling length scale is $l_{\rm cool}\simeq$\,2.5\,kpc in the magnetic arms, where the radiation field is  $U\simeq\, U_{\odot}$ and the magnetic field B$\simeq$~15\,$\mu$G. These lead to a cooling time scale of  $t_{\rm cool}= 1.7 \times 10^{7}$ yr and a diffusion coefficient of $D_0=7.0\times 10^{28}$\,cm$^2$\,s$^{-1}$ for the 2.4\,GeV CREs. 

The above estimates of the CRE diffusion coefficient  agree with both observational  \citep[][]{Strong_98,Dahlem} and theoretical \citep[e.g.][]{Roediger07} estimates of $D_E\simeq(1-10)\times 10^{28}$\,cm$^2$\,s$^{-1}$ and is about 10 times larger than the diffusion coefficient estimated on small scales in the turbulent medium near 30Dor \citep{Murphy_12}. These demonstrate 
the effect of the ISM and CRE cooling on the synchrotron--FIR correlation on kpc-scales. On these (i.e., $\gtrsim$ 1\,kpc) scales, the CRE population is likely to be dominated by old/diffuse particles accelerated in past generations of star
formation.  These particles may even be re-accelerated by passing shocks in the
ISM, and thus will have little memory of their  original birth sites in star-forming regions.   Consequently, the propagation of these particles  is very sensitive to
the ISM conditions.  

On sub-kpc scales, however, the CRE population is more
likely to be dominated by younger CREs, which are still close to their production
sites in star-forming complexes. The energy distribution of these particles
 is more influenced by the age and activity of their acceleration sources rather
than the quasi-state ISM condition. This is in agreement with Murphy et al. (2008)
who showed that the current diffusion length of CREs from star-forming structures
is largely set by the age of the star-formation activity rather than the cooling
mechanisms in the general ISM.

\section{Summary}
Highly resolved and sensitive Herschel images of NGC\,6946 at 70, 100, 160, and 250\,$\mu$m enable us to study the radio--FIR correlation, its variations and dependencies on star formation and ISM properties across the galactic disk.
Our study includes different thermal/nonthermal separation methods. The radio--FIR correlation is calculated using the classical pixel-by-pixel correlation, wavelet scale-by-scale correlation, and the q-method.  The most important findings of this study are summarized as follows:

\begin{itemize}
\item[-]  The slope of the radio-FIR correlation across the galaxy varies as a function of both star formation rate density and magnetic field strength. The total and turbulent magnetic field strengths are correlated with $\Sigma_{\rm SFR}$ with a power-law index of 0.14 and 0.16, respectively (Eqs.14, 15).  This indicates efficient production of turbulent magnetic fields with increasing turbulence in actively starforming regions, in general agreement with \cite{chyzy}.
 
\item[-] In regions where the main heating source of dust is the general ISRF, the synchrotron emission correlates better with the cold dust than with the warm dust. However, there is no difference between the quality of the correlations for colder/warmer dust in regions of a strong radiation field powered by massive stars. This is expected if  warmer dust is mainly heated by SF regions (where synchrotron emission  is produced by young CREs/turbulent magnetic field), and colder dust by a diffuse ISRF across the disk (where synchrotron emission produced by old, diffused CREs/large-scale  magnetic field).  

\item[-] The synchrotron--FIR correlation in strong radiation fields can be well explained by the optically-thin models where massive stars are the common source of the radio and FIR emission. The  intermediate-mass stars seem to be a more appropriate origin for the observed synchrotron--FIR correlation in the ISRF regime.  

\item[-] The synchrotron spectral index map indicates a change in the cooling of CREs when they propagate from their place of birth in star-forming regions across the disk of NGC\,6946. Young CREs emitting synchrotron emission with a flat spectrum, $\alpha_n=0.6\pm0.1$, are found in star-forming regions. Diffused and older CREs (with lower energies) emit synchrotron emission with a steep spectrum, $\alpha_n=1.0\pm0.1$, along the so-called `magnetic arms' (indicating strong synchrotron losses) in the inter-arm regions. The mean synchrotron spectral index is  $\alpha_n=0.8\pm0.1$ across the disk of NGC\,6946. 

\item[-]  The cooling scale length of CREs determined using the multi-scale analysis of the synchrotron--FIR correlation provides an independent measure of the CRE diffusion coefficient. Our determined value of $D_0=4.6\times 10^{28}$\,cm$^2$\,s$^{-1}$ for the 2.2\,GeV CREs agrees with the observed values in the Milky Way. This agreement suggests that, reversing our argument and assuming Milky Way values for $D_E$, the cooling scale length of CREs due to the synchrotron and inverse-Compton energy losses  appear to be consistent with scales on which the radio-FIR correlation is weak on kpc scales. This indicates that the interstellar magnetic fields can affect the propagation of the old/diffuse CREs on large scales.
\end{itemize}
%


\acknowledgement We are grateful to A. Ferguson for kindly providing us with the H$\alpha$ data. The combined molecular and atomic gas data were kindly provided by F. Walter.  FST acknowledges the support by the DFG via the grant TA 801/1-1.



\bibliography{s.bib}  
\appendix
\section{Thermal/nonthermal separation using the standard method}
\cite{Tabatabaei_3_07} presented a detailed comparison of the TRT thermal/nonthermal separation method to a more conventional method, the so called `standard method', for M\,33. The standard method requires accurate absolute measurements of the brightness temperature at high and low radio frequencies and assumes a priori knowledge of $\alpha_n$. An incomplete frequency coverage of the radio images makes it difficult to find variations in  $\alpha_n$, technically, and hence in most cases the thermal/nonthermal components are decomposed assuming a constant $\alpha_n$. Although this assumption does not allow for the study of the change in the CREs energy when they propagate away from SF regions, the standard separation method is often used as it can produce thermal and nonthermal maps using radio continuum data at only two frequencies (at high and low radio band) and more straightforwardly than using, e.g., the TRT method. 
Here,  we derive the thermal/nonthermal distributions from the standard method for NGC\,6946. Comparing the results of the standard and the TRT methods could be instructive for similar studies in future. 

For a total spectral index, $\alpha$, obtained from the observed flux densities at frequencies $\nu_1$ and $\nu_2$ and  constant value of the nonthermal spectral index $\alpha_n$, the thermal fraction at frequency $\nu_1$ is given by 
\begin{equation}
F_{th}^{\nu_1}=((\frac{\nu_2}{\nu_1})^{-\alpha}-(\frac{\nu_2}{\nu_1})^{-\alpha_n})/((\frac{\nu_2}{\nu_1})^{-0.1}-(\frac{\nu_2}{\nu_1})^{-\alpha_n}),
\end{equation}
\citep{Klein_84}. Then the thermal flux density at frequency $\nu_1$, $S_{th}^{\nu_1}$, is obtained from $S_{th}^{\nu_1}=S^{\nu_1} \times F_{th}^{\nu_1}$,
and the nonthermal flux density $S_{n}^{\nu_1}=S^{\nu_1} - S_{th}^{\nu_1}$.
Using the data at 3.5 and 20\,cm for the above formula, and assuming the  frequently adopted index  $\alpha_n\simeq 1$ \citep[e.g. ][]{Klein_84,Berkhuijsen_03}, the corresponding free-free and synchrotron maps are derived \citep[see also ][]{Beck_07}. 
Figure~A.1 shows the resulting maps at 3.5\,cm  divided by those obtained in Sect.~3 (the standard-to-TRT ratio maps).
\begin{figure}
\begin{center}
\resizebox{6cm}{!}{\includegraphics*{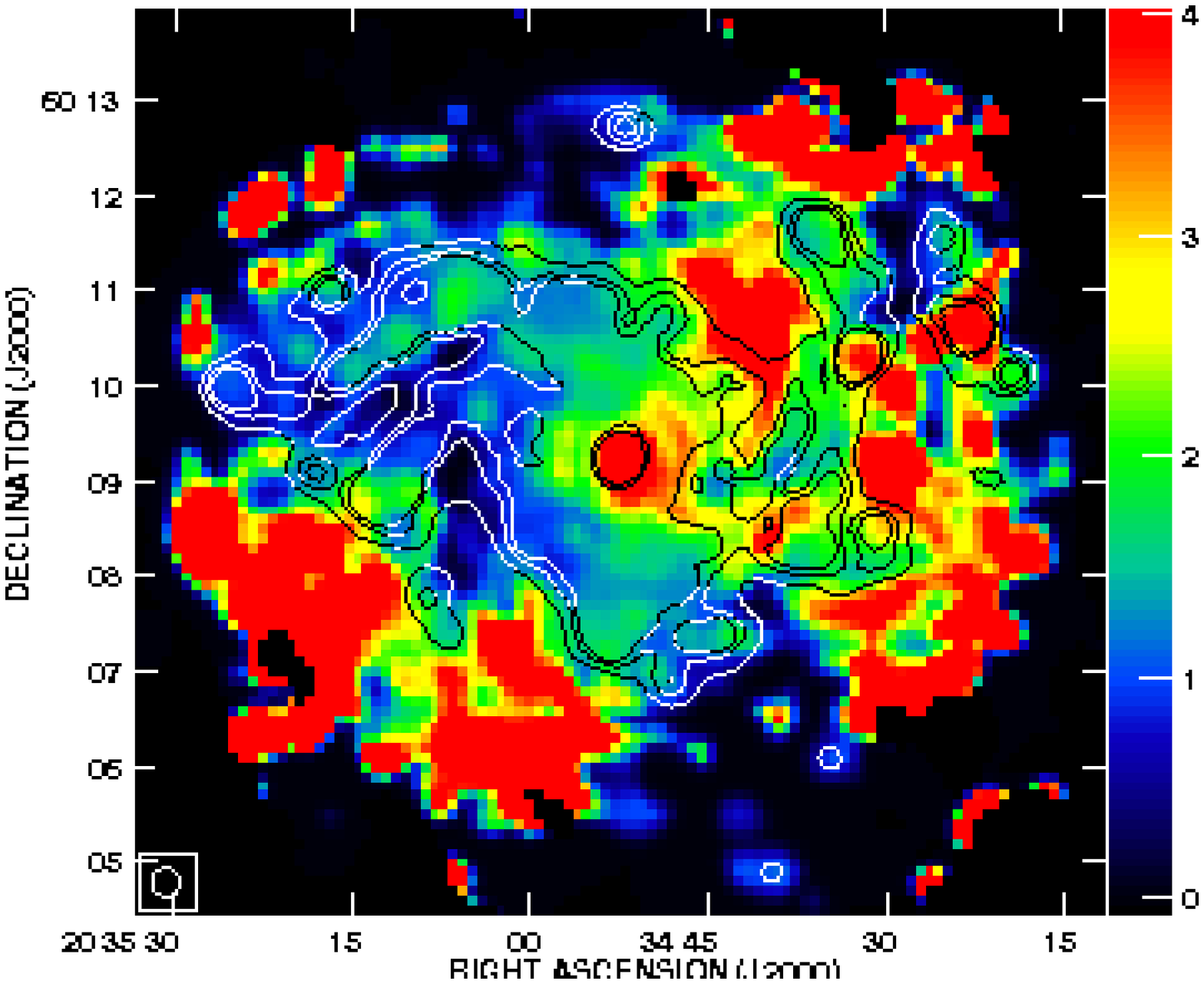}}
\resizebox{6cm}{!}{\includegraphics*{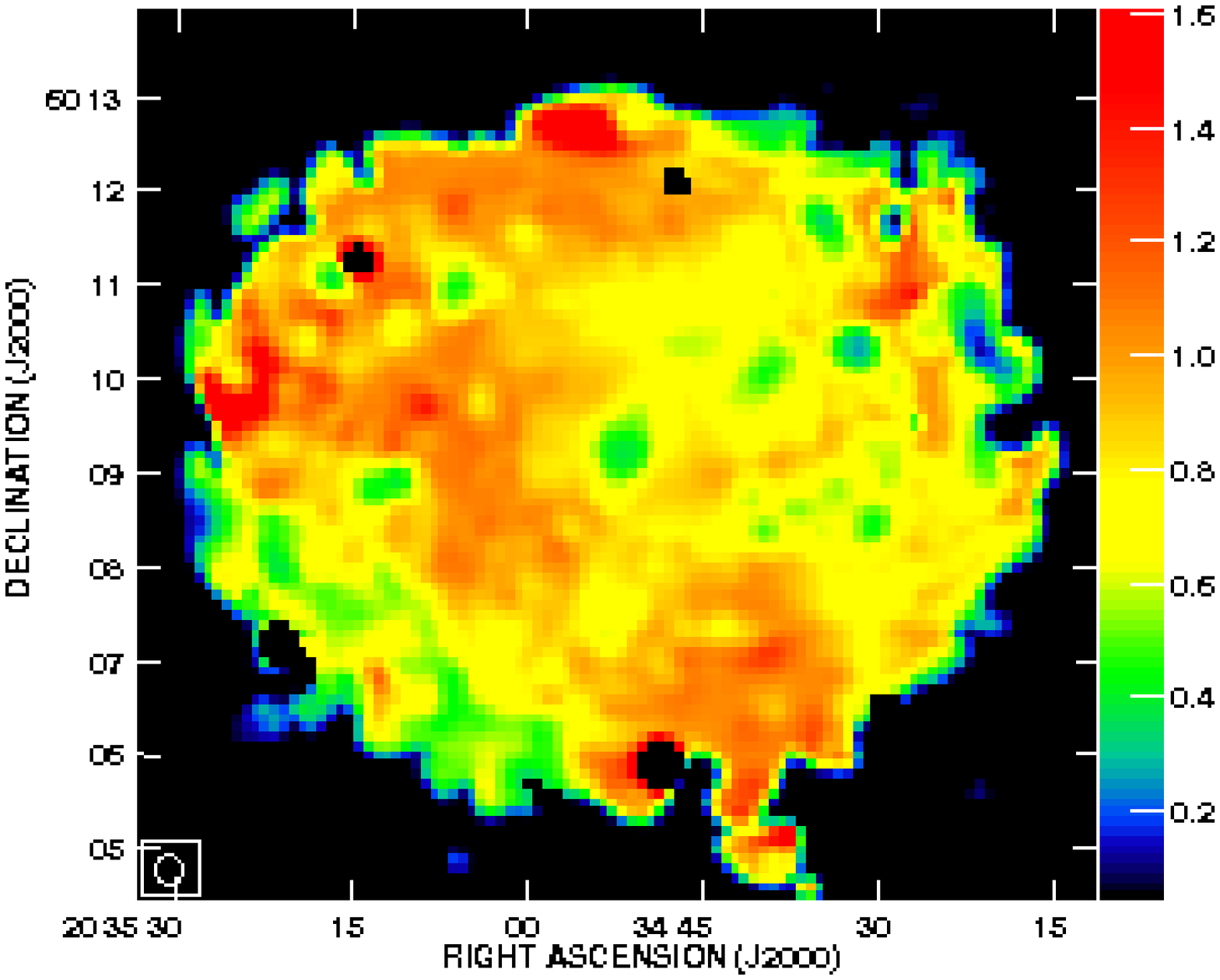}}
\caption[]{\tiny {\it Top:} the ratio map of the standard-to-TRT based free-free intensity overlaid with contours of the TIR emission. The contour levels are 9, 15, 20, 130, 200\,$L_{\odot}$\,kpc$^{-2}$.   {\it Bottom:} the same for the synchrotron component. The ratio values are indicated by the bars at the right of each image.  }
\end{center}
\end{figure}
The standard-to-TRT free-free ratio fluctuates around one 
in the spiral arms and the central disk indicated by the TIR contours. In the nucleus and its surroundings, however, the ratio exceeds 2. A striking excess of free-free emission from the standard method occurs between the arms in the west and the south. We note that, in these regions, there is no significant emission of H$\alpha$ as a tracer of the ionized gas, nor 24 or 70 \,$\mu$m as tracers of warm dust, and even from the TIR as a tracer for SF. Hence, the strong diffuse free-free emission in those regions is not real. 

The standard-to-TRT synchrotron ratio map resembles the $\alpha_n$ map (Fig.~A.1). It is close to unity in regions where   the assumed $\alpha_n$ is close to that obtained from the TRT model. The synchrotron
emission is underestimated in regions with $\alpha_n$ flatter than the assumed $\alpha_n=1$. The lack of  synchrotron emission is particularly seen in regions with an excess of free-free emission in the west/south of the galaxy discussed above. The synchrotron emission from the SF regions and the nucleus is also underestimated as $\alpha_n< 1 $ for those regions (Sect.\,4).

\end{document}